\documentclass[11pt]{article}
\usepackage{verbatim,amsmath,amssymb}
\usepackage{epsfig,float,color}
\usepackage{epstopdf}
\usepackage{geometry}
\usepackage{setspace}
\usepackage{wrapfig}
\usepackage{hyperref}
\usepackage[utf8]{inputenc}
\usepackage{graphicx}
\usepackage{flushend}
\usepackage{tikz,pgfplots}
\usepackage[font={footnotesize}]{caption}
\usepackage[font={footnotesize}]{subcaption}
\usepackage{footnote}
\usepackage{multicol}
\usepackage{multirow}
\usepackage{enumitem}   
\usepackage{algorithm}%
\usepackage{algorithmicx}%
\usepackage{algpseudocode}%
\usepackage{soul}
\usepackage{framed}
\usepackage[titletoc,title]{appendix}
\usepackage{bm}
\usepackage{siunitx}
\usepackage{natbib}
\usepackage[makeroom]{cancel} 
\usepackage{natbib}
\usepackage{hyperref}
\usepackage{bigints} 

\newenvironment{rcases}
{\left.\begin{aligned}}
	{\end{aligned}\right\rbrace}

\definecolor{darkblue}{rgb}{0,0,1}
\hypersetup{pdftex=true, colorlinks=true, breaklinks=true, linkcolor=darkblue, menucolor=darkblue, pagecolor=darkblue, citecolor=darkblue, urlcolor=darkblue}

\definecolor{beaublue}{rgb}{0.94, 0.97, 1.0}
\newcommand{\trr}[1]{{#1}^{\!\top}}

\usepackage{tikz}
\usetikzlibrary{positioning, arrows.meta}
\tikzset{%
	myarrow/.style = {-Stealth, shorten >=5pt}
}

\usepackage{listings}
\definecolor{LightCyan}{rgb}{0.88,1,1}
\definecolor{mygreen}{RGB}{28,172,0} 
\definecolor{mylilas}{RGB}{170,55,241}
\begin{document}
	
	\begin{center}
		\Large{\bf{{\texttt{TOPress3D}: 3D topology optimization with design-dependent pressure loads in MATLAB}}}\\
		
	\end{center}
	
	\begin{center}
		
		\large{Prabhat  Kumar\footnote{\url{pkumar@mae.iith.ac.in}}}
		\vspace{4mm}
		
		\small{\textit{Department of Mechanical and Aerospace Engineering, Indian Institute of Technology Hyderabad, 502285, India}}
		
		\small{\textit{Department of Computational Engineering, Indian Institute of Technology Hyderabad, 502285, India}}\\
		\small{\textit{Department of Engineering Science, Indian Institute of Technology Hyderabad, 502285, India}}\\
		\vspace{4mm}
		 Published\footnote{This pdf is the personal version of an article whose final publication is available at \href{https://link.springer.com/article/10.1007/s11081-024-09931-2}{Optimization and Engineering}}\,\,\,in \textit{Optimization and Engineering}, 
		\href{https://link.springer.com/article/10.1007/s11081-024-09931-2}{DOI:10.1007/s11081-024-09931-2} \\
		Submitted on 11~April 2024, Revised on 26~August 2024, Accepted on 20~September 2024 
	\end{center}
	
	\vspace{1mm}
	\rule{\linewidth}{.15mm}
	{\bf Abstract:}
	This paper introduces ``\texttt{TOPress3D}," a 3D topology optimization  MATLAB code for structures subjected to design-dependent pressure loads. With a primary focus on pedagogical objectives, the code provides an easy learning experience, making it a valuable tool and practical gateway for newcomers, students, and researchers towards this topic. \texttt{TOPress3D} uses Darcy's law with a drainage term to link the given pressure load to design variables that, in turn,  is converted to consistent nodal loads. Optimization problems focused on compliance minimization under volume constraints with pressure loads are solved. Load sensitivities arising due to design-dependent nature of the loads are evaluated using the adjoint-variable approach. The method of moving asymptotes is used to update the design variables. \texttt{TOPress3D} is constituted by six main parts. Each is described in detail. The code is also tailored to solve different problems. The robustness and success of the code are demonstrated in designing a few pressure load-bearing structures. The code is provided in Appendix~\ref{Sec:TOPress3D} and is available with extensions in the supplementary material and publicly at \url{https://github.com/PrabhatIn/TOPress3D}. \\
	

	\vspace{-4mm}
	\rule{\linewidth}{.15mm}

\section{Introduction}
This paper introduces ``TOPress3D," a MATLAB code (158 line) designed for performing 3D topology optimization on structures subjected to design-dependent fluidic pressure loads. While such loads are prevalent in various applications,  addressing them within a topology optimization framework presents distinct challenges as they change  direction, location and/or magnitude with design evolution~\citep{hammer2000topology, kumar2020topology}. These challenges become more pronounced for 3D problems~\citep{kumar2021topology}. Therefore, availability of a publicly accessible pedagogical code can become particularly valuable and can serve as an educational tool and a practical entry point for newcomers, students, and researchers looking to familiarize themselves with this subject. \texttt{TOPress3D} is developed to fill the gap and accomplish the above mentioned objectives.

These days, topology optimization (TO) has become a widely used technique in various applications, as it provides efficient and innovative optimized designs. The technique solves the associated boundary value problems typically using finite element methods, wherein the design domain is discretized by finite elements (FEs), ranging from simple FEs such as triangles, quadrilaterals~\citep{sigmund200199}, and hexahedral FEs~\citep{amir2014multigrid,liu2014efficient} to more advanced ones like honeycomb (hexagonal) tessellations~\citep{saxena2011topology,kumar2022honeytop90}, polygonal elements~\citep{talischi2012polytop,kumar2015topology}, and truncated octahedral FEs~\citep{chi2020virtual,singh2024three}. A design variable $\rho \in [0,\,1]$ is assigned to each element  that determines its state. $\rho=0$ indicates void phase, whereas $\rho=1$ denotes solid phase of the element.  Depending on the nature of the loads, TO approaches can be classified into those with and without design-dependent loads. While many TO approaches exist for the latter, only a few methods have been reported for the former~\citep{picelli2019topology,kumar2020topology}. The number of methods further reduces when considering 3D settings~\citep{kumar2021topology}. One may find 3D approaches only in~\cite{du2004topological,zhang2010topology,yang2005evolutionary,sigmund2007topology,wang2020density,kumar2021topology}. In addition, a 3D paper with related code does not exist yet. Therefore, the current endeavor aims to potentially eliminate barriers hindering the learning and development of 3D TO with design-dependent loads and their extensions for solving different applications experiencing such loads, e.g., pneumatically actuated soft grippers~\citep{pinskier2023automated,pinskier2024diversity}, pressure-loaded meta-materials, to name a few.

Making publicly available education codes in TO is a welcomed trend, which helps the technique grow faster and provides valuable tools. This trend is well accepted in academia and industry, which was started by Sigmund while presenting the first TO educational code having 99-line in MATLAB~\citep{sigmund200199}. Following the trend, many 2D educational papers with codes have been presented for different applications. A list of such codes can be found in~\cite{wang2021comprehensive}. On the other hand, the number of 3D TO papers with code is few, e.g., in~\cite{liu2014efficient,amir2014multigrid,amir2015revisiting,aage2015topology,lagaros2019ac,chi2020virtual,ferrari2020new,schmidt20112589,deng2021efficient,zuo2015simple,fernandez2019aggregation,smith2020matlab,wang2021matlab,du2022efficient,zhao2023matlab,zhuang2023efficient,kim2022matlab}. In addition, one cannot find such a publicly available code for 3D TO with design-dependent pressure loads. The motif herein is to fill the gap, benefiting students, researchers, and practitioners to delve into 3D TO with  design-dependent loads and use and extend the provided code (\texttt{TOPress3D}) for different applications.  

\texttt{TOPress3D} employs 3D hexahedral elements to parameterize design domains. It incorporates the 3D version of the Darcy law, including the drainage term described in~\cite{kumar2021topology}, to establish relationship between the given pressure load and design variables. The 2D counterpart of this relationship is available in~\cite{kumar2020topology} and related MATLAB code in~\cite{kumar2023TOPress}. The variable naming conventions and framework within \texttt{TOPress3D} follow \texttt{TOPress}~\citep{kumar2023TOPress}. The method of moving asymptotes (MMA, cf.~\cite{svanberg1987}) is utilized for updating the design variables in the optimization process. MMA readily permits the code extension with additional physical/geometrical constraints, if any. The assembly process mentioned in~\cite{ferrari2020new} is used for handling the symmetric matrices (stiffness and flow matrices) of~\texttt{TOPress3D}  for efficiency. Whereas, noting that the transformation matrix (Sec.~\ref{Sec:Sec2}) is independent of the design variables, is assembled once before the optimization and used within the optimization process. 

To summarize, this paper offers  TOPress3D MATLAB code with the following new aspects:
\begin{itemize}
		\item The code is developed with 3D hexahedral elements and made publicly available to optimize 3D load-bearing structures with design-dependent pressure loads for researchers, students, and newcomers to this area. It serves as an accessible gateway for extending the code to exciting applications, such as the design of pneumatically activated soft grippers or pressure-loaded metamaterials. In addition, it allows for the inclusion of the advanced constraints, such as buckling and stress, as needed for more complex applications.
		\item  The code employs efficient assembly procedures outlined in~\cite{ferrari2020new} leveraging the symmetry in the elemental stiffness matrix  \texttt{(Ke)} as well as the elemental flow matrices for Darcy  \texttt{(Kp)} and drainage \texttt{(KDp)} using assembly matrices \texttt{iK,\,jK,\,iP,\,jP} (Appendix~\ref{Sec:TOPress3D}). To further reduce memory requirements, the code uses  \texttt{fsparse} function~\citep{engblom2016fast}, which requires \texttt{integer} precision, instead of  \texttt{sparse} function that needs \texttt{double} precision. As a result, the displacement and pressure DOFs are stored as integers (Appendix~\ref{Sec:TOPress3D}). This facilitates solving 3D problems on a laptop with a reasonable  mesh grid.
		\item The code assembles the transformation matrix (Sec.~2) only once before the optimization loop, noticing that the elemental part of it is independent of the design variables using the respective assembly matrices \texttt{iT,\,jT} (Appendix~\ref{Sec:TOPress3D}).
		\item The code contains six main parts, each explained in detail. Its various extensions are also presented.
		\item Code's robustness and efficacy are demonstrated by solving four different design-dependent pressure loadbearing structures.
\end{itemize}

The remainder of the paper is organized as follows. Sec.~\ref{Sec:Sec2} presents topology optimization framework---pressure field evaluation, consistent nodal load calculation, and optimization problem formulation with sensitivity analysis. Sec.~\ref{Sec:Sec3} outlines MATLAB implementation of \texttt{TOPress3D} in detail and provides its various extensions. The numerical results for loadbearing structures are presented in Sec.~\ref{Sec:Sec4}. Lastly, Sec.~\ref{Sec:Sec5} outlines the concluding remarks.

\section{Topology optimization framework}\label{Sec:Sec2}
For completeness, this section briefly outlines pressure load modeling, nodal load evaluation, and nomenclature/parameters' values used in the code. Additionally, it mentions the objective formulation and sensitivity analysis for a compliance problem with a volume constraint. Readers are referred to \cite{kumar2021topology} for a more detailed description.

\subsection{Pressure field evaluation}\label{Sec:pressMod}
According to~\cite{kumar2020topology}, Darcy's law with a drainage term provides an elegant approach to model pressure load in a TO framework. The method has been successfully utilized to solve various problems, including 3D structures and compliant mechanism problems~\citep{kumar2021topology}, length-scale informed pressure-actuated compliant mechanisms~\citep{kumar2022topological}, a PneuNet of a soft robot~\citep{kumar2022towards}, with a featured-based method to obtain close to 0-1 topologies~\citep{kumar2022improved}, multi-material grippers~\citep{pinskier2023automated,pinskier2024diversity}, multi-material frequency-constrained TO with polygonal FEs~\citep{banh2024frequency}, multi-material structures with honeycomb tessellation~\citep{kumar2023topology}, and pneumatically actuated soft robots~\citep{kumar2023sorotop}. The material states of elements change as TO progresses, i.e., one can consider characteristics of  elements like porous media at the beginning with known pressure differences. To this end, Darcy's flux $\mathbf{q}$ is defined as~\citep{kumar2020topology}
\begin{equation}\label{Eq:Darcyflux}
	\bm{q} = -\frac{\kappa}{\mu}\nabla p = -K(\tilde{\bm{\rho}}) \nabla p,
\end{equation}
where   $\nabla p$, $\kappa$, and  $\mu$ indicate the pressure gradient, permeability of the medium, and the fluid viscosity, respectively. $\tilde{\bm{\rho}}$ represents the physical design vector. It is also the filtered design vector \citep{bruns2001} corresponding to the design vector ${\bm{\rho}}$ herein. $K(\tilde{\bm{\rho}})$, flow coefficient, is defined in terms of  $\tilde{\bm{\rho}}$; thus, ${\bm{\rho}}$, to relate the pressure field to the design vector. Mathematically, $K(\tilde{\bm{\rho}})$ for element $e$ is written as
\begin{equation}\label{Eq:Flowcoefficient}
	K(\tilde{\rho_e}) = K_v\left(1-(1-\epsilon) \mathcal{H}(\tilde{{\rho_e}},\,\beta_\kappa,\,\eta_\kappa)\right),
\end{equation} 
where $\epsilon = \frac{K_s}{K_v}$ represents the flow contrast.  $K_v$ and $K_s$ indicate the flow coefficients of  void and solid phases of an element, respectively, and
\begin{equation}\label{Eq:SmoothHev}
	\mathcal{H}(\tilde{{\rho_e}},\,\beta_\kappa,\,\eta_\kappa) = \frac{\tanh{\left(\beta_\kappa\eta_\kappa\right)}+\tanh{\left(\beta_\kappa(\tilde{\rho}_e - \eta_\kappa)\right)}}{\tanh{\left(\beta_\kappa \eta_\kappa\right)}+\tanh{\left(\beta_\kappa(1 - \eta_\kappa)\right)}}
\end{equation}
 is a smooth Heaviside function. We write  $\{\eta_\kappa,\,\beta_\kappa\}$ the flow parameters~\citep{kumar2023TOPress}, where $\eta_\kappa$ and $\beta_\kappa$ indicate the step position and slope of $K(\tilde{\rho_e})$, respectively. We set  $K_v =1$, and $\epsilon = \SI{1e-7}{}$, i.e., $K_s = \epsilon$ in \texttt{TOPress3D} code.  $K(\tilde{\rho_e})$ is defined as~\citep{kumar2020topology}
 \begin{equation}
 	K(\tilde{\rho_e}) = 1-(1-\epsilon) \mathcal{H}(\tilde{{\rho_e}},\,\beta_\kappa,\,\eta_\kappa).
 \end{equation}

The equilibrium equation corresponding to~Eq.~\ref{Eq:Darcyflux} is~\citep{kumar2020topology}
\begin{equation}\label{Eq:DarcyFEM}
	\nabla \cdot \bm{q} = -\nabla \cdot \left(K(\tilde{\bm{\rho}}) \nabla p\right) = 0.
\end{equation}
\begin{figure}[h]
	\centering
	\includegraphics[scale = 1]{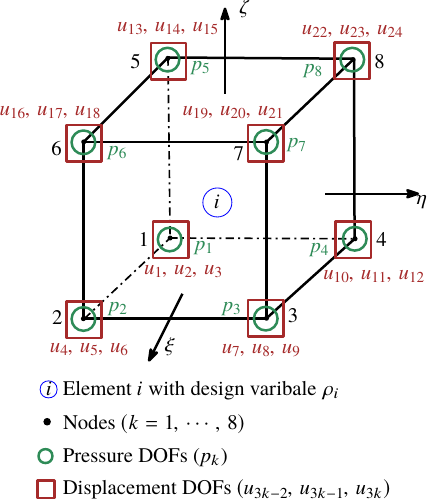}
	\caption{Element $i$ nomenclature. \texttt{Lface}, \texttt{Rface}, \texttt{BTface}, \texttt{Tface}, \texttt{Fface} and \texttt{Bface} represent left, right, bottom, top, front, and back faces, respectively. Herein, these faces respectively contain $\left\{1,\,2,\,5,\,6\right\}$, $\left\{3,\,4,\,7,\,8\right\}$, $\left\{1,\,2,\,3,\,4\right\}$, $\left\{5,\,6,\,7,\,8\right\}$, $\left\{2,\,3,\,6,\,7\right\}$ and $\left\{1,\,4,\,5,\,8\right\}$ nodes. }
	\label{fig:DOFs}
\end{figure}
A drainage term is included in Darcy's law (Eq.~\ref{Eq:Darcyflux}) to achieve a realistic pressure field~\citep{kumar2021topology}. As a result, Eq.~\ref{Eq:DarcyFEM} transpires to
\begin{equation}\label{Eq:stateequation}
	\nabla\cdot\bm{q} - Q_\text{drain} = \nabla \cdot \left(K(\tilde{\bm{\rho}}) \nabla p\right)+  Q_\text{drain}=0,
\end{equation}
with ${Q}_\text{drain}= -D(\bar{\rho_e}) (p - p_{\text{ext}})$, where $D(\bar{\rho_e}) =  \linebreak D_{\text{s}}\mathcal{H}(\bar{{\rho_e}},\,\beta_d,\,\eta_d)$. $\left\{\eta_\text{d},\,\beta_\text{d}\right\}$, termed the drainage parameters, are analogous to $\{\eta_\kappa,\,\beta_\kappa\}$. To restrict the user-defined parameters, \texttt{TOPress3D} code considers $\eta_d = \eta_k = \eta_f$ and $\beta_d = \beta_k =\beta_f$. In addition, users are free to select these parameters as per the recommendation provided in~\cite{kumar2020topology}. 
\begin{equation}\label{Eq:Ds}
	D_{\text{s}} = \left(\frac{\ln{r}}{\Delta s}\right)^2 K_\text{s},\,\text{with}\,\, r = \frac{p|_{\Delta s}}{p_\text{in}},
\end{equation}
where $\Delta s$, a penetration parameter, is set equal to size of a few FEs. $p|_{\Delta s}$ indicates the pressure at $\Delta s$. One can use  $r\in[0.001 \,\,  0.1]$~\citep{kumar2023TOPress}.

As mentioned, the code uses hexahedral elements (Fig.~\ref{fig:DOFs}) to describe the design domain. A hexahedral element with node numbers is displayed in Fig.~\ref{fig:DOFs}. With standard finite element  method (FEM), Eq.~\ref{Eq:stateequation} is written as~\citep{kumar2021topology}
\begin{equation} \label{Eq:FEAformulation}
	\begin{aligned}
		&\left[\int_{\mathrm{\Omega}_e}\left( K~ \mathbf{B}^\top_\text{p} \mathbf{B}_\text{p}\right) \text{d} V  + \int_{\mathrm{\Omega}_e}\left( D ~\mathbf{N}^\top_\text{p} \mathbf{N}_\text{p}\right) \text{d} V~\right]\mathbf{p}_e = \mathbf{0},
	\end{aligned}
\end{equation}
while considering $p_\text{out} = 0$ and $\mathbf{q}_\mathrm{\Gamma} =0$. $\mathbf{N}_\text{p} = [N_1,\,N_2,\,N_3,\,\cdots,\,N_8]$ are the trilinear shape functions (Appendix~\ref{Sec:APPExpression}) for the hexahedral elements (Fig.~\ref{fig:DOFs}), $\mathbf{B}^\top_\text{p} = \nabla\mathbf{N}_\text{p}$, and $\mathbf{p}_e = [p_1,\,p_2,\,p_3,\,\cdots,\,p_8]^\top$. Note that $N_k$ and $p_k$ ($k = 1,\,\cdots,\,8$) indicate shape function and pressure degree of freedom for the $k^\text{th}$ node of hexahedral element~$i$, respectively (Fig.~\ref{fig:DOFs}).
To this end, Eq.~\eqref{Eq:FEAformulation} yields to
\begin{equation} \label{Eq:FEAnumint}
	\begin{aligned}
		\mathbf{K}_\text{p}^e \mathbf{p}_e+ \mathbf{K}^e_\text{Dp} \mathbf{p}_e = \mathbf{A}_e \mathbf{p}_e = \mathbf{0},
	\end{aligned}
\end{equation}
where  element flow matrices of the Darcy and drainage parts are $\mathbf{\mathbf{K}}_\text{p}^e$ and $\mathbf{K}^e_\text{Dp}$, respectively. They are evaluated and their numerical value are provided in Appendix~\ref{Sec:APPExpression}. $\mathbf{A}_e = \mathbf{\mathbf{K}}_\text{p}^e + \mathbf{K}^e_\text{Dp}$ is the overall element flow matrix. Upon assembly, Eq.~\ref{Eq:FEAnumint} yields to
\begin{equation}\label{Eq:PDEsolutionpressure}
	\mathbf{Ap} = \mathbf{0}.
\end{equation}
Since  $\mathbf{\mathbf{K}}_\text{p}^e$ and $\mathbf{K}^e_\text{Dp}$ are  symmetric matrices; thus, $\mathbf{A}$ is also a symmetric matrix. Eq.~\ref{Eq:PDEsolutionpressure} is solved while applying the given pressure load boundary conditions. $\mathbf{A}$ and $\mathbf{p}$ are sub-blocked  into free and  prescribed pressure DOFs, denoted by subscripts $f$ and $p$, respectively; Eq.~\ref{Eq:PDEsolutionpressure} is rewritten as~\citep{kumar2023sorotop}
\begin{equation}\label{Eq:partEq6}
	\begin{bmatrix}
		\mathbf{A}_{ff} & \mathbf{A}_{fp} \\
		\mathbf{A}_{fp}^\top & \mathbf{A}_{pp} 
	\end{bmatrix}  
	\begin{bmatrix}
		\mathbf{p}_f \\
		\mathbf{p}_p 
	\end{bmatrix} = \begin{bmatrix}
		\mathbf{0} \\
		\mathbf{0}
	\end{bmatrix}
\end{equation}
Eq.~\ref{Eq:partEq6} results in the following equations for free and prescribed pressure DOFs
	\begin{equation}\label{Eq:partEq7}
		\mathbf{A}_{ff}\mathbf{p}_f + \mathbf{A}_{fp}\mathbf{p}_p = \mathbf{0},
	\end{equation}
	and
	\begin{equation}\label{Eq:partEq8}
		\mathbf{A}^T_{fp}\mathbf{p}_f + \mathbf{A}_{pp}\mathbf{p}_p = \mathbf{0},
	\end{equation}
	respectively. Since $\mathbf{p}_p$ is known, $\mathbf{p}_f$ is determined using Eq.~\ref{Eq:partEq7} yielding $\mathbf{p}_f = \mathbf{A}_{ff}^{-1}\mathbf{A}_{fp}\mathbf{p}_p$. In \texttt{TOPress3D } code, we use the MATLAB \texttt{decomposition} function (line~123) to perform the above operations and compute $\mathbf{p}_f$. To verify the accuracy of $\mathbf{p}_f$, one can substitute it back into Eq.~\ref{Eq:partEq8}, which is typically necessary when using an indirect method to solve Eq.~\ref{Eq:partEq7} for determining $\mathbf{p}_f$. Finally, one gets the pressure field $\mathbf{p}$ as a function of physical variables (design variables) as $\mathbf{p}_f$ is now known.
\subsection{Nodal load evaluation}
The pressure field $\mathbf{p}$ is transformed into nodal loads by applying equilibrium conditions on an elemental cube experiencing body forces~\citep{kumar2020topology,kumar2021topology}. We get~\citep{kumar2020topology}
\begin{equation}\label{Eq:pressuretoload}
	\bm{b} \text{d}V = -\nabla p \text{d}V,
\end{equation}
where $\text{d}V$ and $\bm{b}$ indicate elemental volume and body force per unit volume, respectively. Given standard FEM, the nodal force $\mathbf{F}_e $ for an element is determined  as~\citep{kumar2021topology}
\begin{equation}\label{Eq:Forcepressureconversion}
	\begin{split}
		\mathbf{F}_e &= \int_{\mathrm{\Omega}_e} \trr{\mathbf{N}}_\mathbf{u}\bm{b}  d {V} =  -\int_{\mathrm{\Omega}_e} \trr{\mathbf{N}}_\mathbf{u}\nabla p  d {V}\\&= -\left[\int_{\mathrm{\Omega}_e} \trr{\mathbf{N}}_\mathbf{u} \mathbf{B}_p  d {V}\right]\, \mathbf{p}_e = \mathbf{T}_e\,\mathbf{p}_e,
	\end{split}
\end{equation}
where $\mathbf{N}_\mathbf{u} = [N_1\mathbf{I},\, N_2\mathbf{I},\,N_3\mathbf{I},\cdots,\,\,N_8\mathbf{I}]$, with $\mathbf{I}$  the identity matrix in $\mathcal{R}^3$. $\mathbf{F}_e $ is assembled to achieve its global nodal force vector 	$\mathbf{F}$. Expression and numerical value for $\mathbf{T}_e$ are provided in Appendix~\ref{Sec:APPExpression}. Eq.~\ref{Eq:pressuretoload} transpires to 
\begin{equation}\label{Eq:nodalforce}
	\mathbf{F} = -\mathbf{T}\mathbf{p},
\end{equation}
in the global sense. $\mathbf{T}$ is the global transformation matrix. Note that $\mathbf{T}_e$ is independent of design variables; thus, $\mathbf{T}$ is assembled prior to the optimization steps. This step helps save computational requirements; thus, making code relatively efficient. 
\subsection{Objective and sensitivity analysis}\label{Sec:TopOptFor}
We solve the following optimization problem for the three-dimensional loadbearing structures with a given volume constraint:
\begin{equation} \label{Eq:OPTI} 
	\begin{rcases}
		\begin{split}
			&{\min_{\tilde{\bm{\rho}}}} \quad C({\tilde{\bm{\rho}}}) = \mathbf{u}^\top \mathbf{K}(\tilde{\bm{\rho}})\mathbf{u} = \sum_{j=1}^{{nel}}\mathbf{u}_j^\top\mathbf{k}_j(\tilde{\rho}_j)\mathbf{u}_j\\
			&\text{subjected to:}\\
			&\qquad\qquad \qquad\bm{\lambda}_1:\,\,\mathbf{A} \mathbf{p} = \mathbf{0}\\
			&\qquad \qquad\qquad\bm{\lambda}_2:\,\,\mathbf{K} \mathbf{u} = \mathbf{F} = -\mathbf{T}\mathbf{p}\\
			&\qquad \qquad\qquad\Lambda:V(\tilde{\bm{\rho}})-V^* \le 0\\
			&\qquad \qquad\qquad\quad\,\,\,\, \bm{0} \leq \tilde{\bm{\rho}} \leq \bm{1} \\
			&\text{Data:} \quad V^*,\,E_0,\,E_\text{min},\,p,\,K_v,\,\epsilon,\,\eta_f,\,\beta_f
		\end{split}
	\end{rcases},
\end{equation} 
where $C$ and $nel$ denote the structure's compliance and total number of elements used to represent design domain ($\mathrm{\Omega}_e$). $\mathbf{K}$ is the global stiffness matrix, whereas $\mathbf{u}$ indicates the global displacement vector. $\mathbf{k}_j$ and $\mathbf{u}_j$ are  stiffness matrix and  displacement vector for element~$j$, respectively. $V^*$ and $V$ denote the permitted and current volume of $\mathrm{\Omega}_e$, respectively. $\mathbf{F}$ indicates the global force vector. $\bm{\lambda}_1$, $\bm{\lambda}_2$  and $\Lambda$ are the Lagrange multipliers. The former two are vectors, whereas the latter one is a scalar. Vector $\tilde{\bm{\rho}}$ represent filtered counterparts of design variable vector  $\bm{\rho}$. The first vector is termed the physical vector herein.

One determines  filtered design variable for element $i$ as:
\begin{equation} \label{EQ:density_filter}
	\tilde{\rho}_i = \frac{\sum_{j=1}^{{nel}}\rho_j \mathrm{v}_j \mathrm{w}(\mathbf{x}_i,\mathbf{x}_j)}{\sum_{j=1}^{{nel}} \mathrm{v}_j \mathrm{w}(\mathbf{x}_i,\mathbf{x}_j) } \; ,
\end{equation}
where $\mathrm{w}(\mathbf{x}_i,\mathbf{x}_j)$ = $\mathrm{max}  \left(0 \; , \; 1-\frac{\| \mathrm{\mathbf{x}}_i - \mathrm{\mathbf{x}}_j \|}{\mathrm{r}_\mathrm{fill}} \right)$ \citep{bruns2001}, $r_\text{fill}$ and $\mathrm{v}_j$ are filter radius and volume of element $j$, respectively. One can determine $\mathrm{v}_j$ and $\mathrm{w}(\mathrm{x}_i,\mathrm{x}_j)$ prior to the optimization and can store in a matrix $\mathbf{H}$ as:
\begin{equation}
	\mathrm{H}_{i,j} = \frac{\mathrm{v}_j \: \mathrm{w}(\mathrm{\mathbf{x}}_i,\mathrm{\mathbf{x}}_j)}{{\sum_{j=1}^{\texttt{nel}} \mathrm{v}_j \mathrm{w}(\mathrm{\mathbf{x}}_k,\mathrm{\mathbf{x}}_j) }} \: .
\end{equation} 
The filtered design vector and its derivative with respect to the design vector can be written as $\bm{\tilde{\rho}}=\mathbf{H}\bm{\rho}$  and $\frac{\partial \bm{\tilde{\rho}}}{\partial \bm{\rho}}=\mathbf{H}^\top$, respectively. $\texttt{TOPress3D}$ uses \texttt{imfilter} MATLAB function for the filtering operations (see Appendix~\ref{Sec:TOPress3D}). 

The modified Solid Isotropic Material with Penalization (SIMP) interpolation scheme is used.
Young's modulus of element $i$, $E_i$, is written as
\begin{equation}\label{Eq:SIMPformulation}
	\mathrm{E}_i = \mathrm{E}_\mathrm{min} + \tilde{\rho}^p_i(\mathrm{E}_1-\mathrm{E}_\mathrm{min}),
\end{equation}
where $p$ is the SIMP parameter. $E_1$ and $E_\text{min}$ are Young's moduli of element's solid and void states, respectively.

\subsubsection{Sensitivity analysis}
We use the method of moving asymptotes (MMA, cf.~\cite{svanberg1987}), a gradient-based optimizer, for updating the design variables. Thus, we need objective's and constraint's derivatives with respect to the design variables, which are determined using the adjoint-variable method herein. The augmented performance function $\mathcal{L}$ in terms of  objective function and equilibrium equations~\eqref{Eq:OPTI} can be written as~\citep{kumar2023TOPress}

\begin{equation}\label{Eq:Lagrange}
	\mathcal{L} = \mathbf{u}^\top \mathbf{K}\mathbf{u} + \bm{\lambda}_1^\top \mathbf{AP} + \bm{\lambda}_2^\top \left(\mathbf{KU + TP}\right).
\end{equation} 
Equation~\eqref{Eq:Lagrange} is differentiated with respect to the physical design variable, and rearranging, one gets 
\begin{equation}\label{Eq:Lagrangeder}
	\begin{split}
		\frac{d \mathcal{L}}{d \tilde{{\rho}}_i} 
		=& \mathbf{u}^\top \frac{\partial\mathbf{K}}{\partial \tilde{{\rho}}_i}\mathbf{u} + \bm{\lambda}_2^\top \left(\frac{\partial\mathbf{K}}{\partial \tilde{{\rho}}_i}\mathbf{u}\right) + \bm{\lambda}_1^\top \left(\frac{\partial\mathbf{A}}{\partial\tilde{{\rho}}_i}\mathbf{p}\right) \\ &+ \underbrace{\left(2\mathbf{u}^\top\mathbf{K} + \bm{\lambda}_2^\top \mathbf{K}\right)}_{\Xi_1}\frac{\partial\mathbf{u}}{\partial \tilde{{\rho}}_i} + \underbrace{\left(\bm{\lambda}_1^\top\mathbf{A} + \bm{\lambda}_2^\top \mathbf{T} \right)}_{\Xi_2}\frac{\partial\mathbf{p}}{\partial\tilde{{\rho}}_i}
	\end{split}
\end{equation} 
One use $\Xi_1 = 0$ and $\Xi_2 = 0$ to determine  $\bm{\lambda}_1$ and  $\bm{\lambda}_2$ from the above equation, i.e.,
\begin{equation} \label{Eq:LagrangeMultiplier}
	\begin{aligned}
		\bm{\lambda}_2 &= -2\mathbf{u}, \\
		\bm{\lambda}_1^\top &= - 	\bm{\lambda}_2^\top \mathbf{T} \mathbf{A}^{-1} = 2\mathbf{u}^\top \mathbf{T} \mathbf{A}^{-1},
	\end{aligned}
\end{equation}
and, therefore,
\begin{equation}\label{Eq:senstivities_Obj}
	\frac{d {C}}{d \tilde{{\rho}}_i} = -\mathbf{u}^\top \frac{\partial\mathbf{K}}{\partial \tilde{{\rho}}_i}\mathbf{u} + \underbrace{2\mathbf{u}^\top \mathbf{T} \mathbf{A}^{-1}\frac{\partial\mathbf{A}}{\partial\tilde{{\rho}}_i}\mathbf{p}}_{\text{Load sensitivities}}
\end{equation}
Load sensitivities, appeared in Eq.~\eqref{Eq:senstivities_Obj} due to the design-dependent nature of the load, affect the optimized topologies as demonstrated in prior work such as~\cite{kumar2023sorotop,kumar2020topology,kumar2023TOPress}. Therefore, neglecting them during optimization may not be advisable. Finally, using the chain rule, the objective derivatives can be determined~\citep{kumar2023TOPress}. Calculation of the derivative of the volume constraint is straightforward~\citep{sigmund200199}. In the subsequent section, we provide the MATLAB implementation for \texttt{TOPress3D} code.  

\section{Structure of \texttt{TOPress3D}}\label{Sec:Sec3}
The section provides a complete description of the MATLAB code, \texttt{TOPress3D}. Reader can download the code, provided in Appendix~\ref{Sec:TOPress3D}, and its extensions from the supplementary material of the paper. One calls the code in the MATLAB command window as
\begin{lstlisting}[basicstyle=\footnotesize\ttfamily,breaklines=true,numbers=none,frame = tb, backgroundcolor=\color{beaublue}]
TOPress3D(nelx,nely,nelz,volf,penal,rmin,etaf,betaf,lst,maxit)}
\end{lstlisting}

\noindent where \texttt{nelx}, \texttt{nely} and \texttt{nelz} indicate the number of elements in $x-$, $y-$ and $z-$directions, respectively. \texttt{volf} represents the given volume fraction, \texttt{penal} refers to the penalty parameter of the SIMP technique (Eq.~\ref{Eq:SIMPformulation}), \texttt{rmin} is the filter radius, \texttt{etaf} and \texttt{betaf} are related to the flow coefficient and drainage term, respectively. \texttt{lst} indicates the status of load-sensitivities. \texttt{lst} = 1 indicates that load sensitivities are considered in the optimization process, whereas \texttt{lst} = 0 means otherwise. \texttt{maxit} variable indicates the maximum number of MMA iterations. Hexahedral elements are used for discretizing the domains. Local degree of freedoms (DOFs) pertaining to displacement and pressure are shown in Fig.~\ref{fig:DOFs}. \texttt{TOPress3D} contains the following six main parts:

\begin{enumerate}
		\item[(I)] MATERIAL AND FLOW PARAMETERS INITIALIZATION
		\item[(II)] FINITE ELEMENT ANALYSIS AND PASSIVE  SOLID/VOID REGIONS PREPARATION
		\item[(III)] ASSIGNING PRESSURE B.Cs, DISPLACEMENT B.Cs, AND LAGRANGE MULTIPLIERS INITIALIZATION
		\item[(IV)] FILTER PREPARATION
		\item[(V)] MMA OPTIMIZATION PREPARATION AND INITIALIZATION
		\item[(VI)] MMA OPTIMIZATION LOOP
		\begin{itemize}
			\item [(VI.1)] SOLVING FLOW BALANCED EQUATION
			\item [(VI.2)] DETERMINING CONSISTENT NODAL LOADS AND GLOBAL DISPLACEMENT VECTOR
			\item [(VI.3)] OBJECTIVE, CONSTRAINT AND THEIR SENSITIVITIES COMPUTATION
			\item [(VI.4)] SETTING AND CALLING MMA OPTIMIZATION
			\item [(VI.5)] PRINTING AND PLOTTING RESULTS
		\end{itemize}
\end{enumerate}
We describe each part in detail below:

\noindent (I) \textbf{MATERIAL AND FLOW PARAMETERS INITIALIZATION:} \texttt{E1} (line~3) and \texttt{Emin} (line~4) indicate  $E_1$ (Eq.~\ref{Eq:SIMPformulation})  and $E_\text{min}$ (Eq.~\ref{Eq:SIMPformulation}), respectively. Line~5 mentions Poisson's ratio, \texttt{nu}, which is set to 0.30. On line~6, values of $K_v$ (Eq.~\ref{Eq:Flowcoefficient}), indicated by \texttt{Kv}, $\epsilon$ (Eq.~\ref{Eq:Flowcoefficient}), indicated by \texttt{epsf},  $r$ and $\Delta s$, indicated by \texttt{Dels} (Eq.~\ref{Eq:Ds}), are given utilizing \texttt{deal} MATLAB function.  \texttt{deal} function creates multiple output variables with specified values. $D_s$ and $(K_v-K_s)$ are denoted by \texttt{Ds} and \texttt{Kvs}, respectively, and are determined on line~7. 

\noindent (II) \textbf{FINITE ELEMENT ANALYSIS AND PASSIVE  SOLID/VOID REGIONS PREPARATION:} This part provides FE analysis preparation for flow and structure parts of the code on lines 8-87. The part at end also facilitates the inclusion of passive solid/void regions, if any. Number of nodes in $x-$, $y-$ and $z-$directions are recorded in \texttt{ndx}, \texttt{ndy} and \texttt{ndz}, respectively on line~9. \texttt{nel} and \texttt{nno} indicate the total number of FEs and nodes, respectively.  \texttt{nel} and \texttt{nno} are determined on line~10. As node numbers and associated displacement and pressure DOFs are integers, we use \texttt{int32} MATLAB function while recording them. Instead of using \texttt{sparse} MATLAB function, we use \texttt{fsparse} routine, created by~\citep{engblom2016fast} to perform the assembly procedure. The former records locations (DOFs-rows and columns) information as double precision numbers, whereas the latter records them as integers, thus saving computational requirements that, in turn, make the procedure computationally efficient~\citep{ferrari2020new}. Next, we mention the procedure to use/install \texttt{fsparse}.

 For using \texttt{fsparse}, one can download the ``\texttt{stenglib}" library\footnote{\url{https://github.com/stefanengblom/stenglib}} and install it as per the \texttt{README.md} file. Extract the downloaded `stenglib-master.zip' file, copy the folder ``Fast" and make it the current folder in  MATLAB and then type `make' in the command windows and press `Enter' bottom on the keyboard. If MATLAB requests to install `MinGW64 Compiler (C)', download and install the said compiler and run `make' again as procedure said earlier. Once the compilation is done, the user can see the `MEX-file' in the ``Fast'' folder. One can then place \texttt{TOPress3D.m} code and its extensions with \texttt{mmasub.m} and \texttt{subsolv.m} files in `Fast' folder and execute \texttt{TOPress3D.m} and its extensions as suggested in Sec.~\ref{Sec:Sec4}.

The matrix containing displacement DOFs, recorded in \texttt{Udofs}, is created on lines~13-14 using array \texttt{nodenrs} (line~11) and vector \texttt{edofVec} (line~12). Line~15 determines pressure DOFs, all pressure DOFs, and displacement DOFs. They are recorded in \texttt{Pdofs}, \texttt{allPdofs} and \texttt{allUdofs}, respectively. Nodes constituting faces are determined next, as they are required to apply the given pressure loads. In that view, lines~16-18 determine nodes that form the bottom and top faces of domain in vectors~\texttt{BTface} and \texttt{Tface}, respectively. In addition, nodes making the left and right faces are recorded on line~19 in vectors \texttt{Lface} and \texttt{Rface}, respectively. Further, line~20 determines nodes constituting the front and back faces in vectors \texttt{Fface} and \texttt{Bface}, respectively. Vectors \texttt{iK} and \texttt{jK}, required for performing assembly of the stiffness matrix, are determined on line~26 as per~\cite{ferrari2020new}. To reduce the assembly indexing, \texttt{Iar} is determined on line~27, which is used on line~127. Lines~28-51 record the lower half of the elemental stiffness matrix in  vector~\texttt{Ke}.  \texttt{Ke} is used for the assembling the stiffness matrix on line~127; also, to recover the complete elemental matrix \texttt{Ke0} on lines~52. \texttt{Ke0} (lines~52-54) is used to evaluate the compliance sensitivity on line~133. Following the above steps, vectors \texttt{iP} and \texttt{jP}, analogous to vectors \texttt{iK} and \texttt{jK}, are determined on line~59 for flow matrix assembly.  \texttt{IarP} analogous to \texttt{Iar} is determined on line~60. Element flow matrix corresponding to the Darcy law (\texttt{Kpl}) and drainage term (\texttt{KDpl}), in the factorized form, are recorded on lines~61-62 and lines~63-64, respectively. \texttt{Kpl} and \texttt{KDpl} indicate $\mathbf{K}_\text{p}^e$ and $\mathbf{K}_\text{DP}^e$ (Eq.~\ref{Eq:FEAnumint}), respectively with unit $K$ (Eq.~\ref{Eq:FEAformulation}) and $D$ (Eq.~\ref{Eq:FEAformulation}) (see Appendix~\ref{Sec:APPExpression}). The corresponding full matrices \texttt{Kp} and \texttt{KDp} are recovered between lines~65-67. These matrices are needed while determining the load sensitivities (lines~134-136). Lengths of vectors \texttt{Ke} and \texttt{Kpl} are determined on line~68 utilizing, which are used later on lines~126 and line~117, respectively. On lines~69-75, the code records elemental transformation matrix $\mathbf{T_e}$ (Eq.~\ref{Eq:Forcepressureconversion}) in the vectorized form in vector \texttt{Te}. As \texttt{Te} is a rectangular matrix, the assembly procedure to determine the global transformation matrix $\mathbf{T}$ (Eq.~\ref{Eq:nodalforce})  is similar to~\texttt{TOPress} code. Vectors \texttt{iT} and \texttt{jT} are created on liens~76 and 77, respectively. Noting that the transformation matrix is independent of the design variables, the global form $\mathbf{T}$ (Eq.~\ref{Eq:nodalforce}) is determined on line~79 and recorded in \texttt{TG}. This steps save computational requirements.

Smooth Heaviside projection function (Eq.~\ref{Eq:SmoothHev}), \texttt{IFprj}, needed to define the flow (Eq.~\ref{Eq:Flowcoefficient}) and drainage (Eq.~\ref{Eq:stateequation}) coefficients, is defined on lines~80-81. The function has three input variables. Lines~82-83 determine its derivative with respect to the first variable, i.e., the design variable, and recorded in a function~\texttt{dIFprj}. The derivative function is required on lines~134-135 to determine the load sensitivities. Line~84 gives room to include passive solid (\texttt{NDS})/void (\texttt{NDV}) regions, if any. Active design vector is determined on line~85 and is recorded in \texttt{act}.

\noindent (III) \textbf{ASSIGNING PRESSURE B.Cs, DISPLACEMENT B.Cs, AND LAGRANGE MULTIPLIERS INITIALIZATION:} This part of the code assigns pressure load and boundary conditions, displacement boundary conditions (fixed and free DOFs). This part also initializes the global displacement and $\bm{\lambda}_1$ (Eq.~\ref{Eq:LagrangeMultiplier}). On line~87, vector \texttt{PF}  initializes pressure load vector, and scalar \texttt{Pin} contains the magnitude of the input pressure load. Line~88 updates \texttt{PF} as per the applied pressure loading locations. Fixed pressure DOFs, recorded in\linebreak \texttt{fixedPdofs}, and free pressure DOFs, stored in \texttt{freePdofs}, are determined on line~89 and line~90, respectively. On line~91, array \texttt{pfixeddofsv} records the fixed pressure DOFs and corresponding values in its first and second columns, respectively. Lines~92-93 record fixed displacement nodes in vector \texttt{fixnn}. Vector \texttt{fixedUdofs} stores fixed displacement DOFs (line~94). Free displacement DOFs are recorded in \texttt{freeUdofs}  (line~95). The global displacement vector \texttt{U} and Lagrange multiplier \texttt{lam1} for the sensitive analysis are initialized on line~96.    

\noindent (IV) \textbf{FILTER PREPARATION:}  \texttt{imfilter} MATLAB function is used for performing the density filtering. The filter variable \texttt{Hs} is determined on lines~97-101. One can also determine the filter parameters as~\citep{amir2014multigrid,amir2015revisiting,liu2014efficient}:
\begin{lstlisting}[basicstyle=\footnotesize\ttfamily,breaklines=true,numbers=none,frame=tb,backgroundcolor=\color{beaublue}]
iH = ones(nelx*nely*nelz*(2*(ceil(rmin)-1)+1)^3,1);
jH = ones(size(iH));
sH = zeros(size(iH));
k = 0;
for i1 = 1:nelx
    for k1 = 1:nelz
        for j1 = 1:nely
            e1 = (i1-1)*nely*nelz + (k1-1)*nely + j1;
            for i2 = max(i1-(ceil(rmin)-1),1):min(i1+(ceil(rmin)-1),nelx)
                for k2 = max(k1-(ceil(rmin)-1),1):min(k1+(ceil(rmin)-1),nelz)
                    for j2 = max(j1-(ceil(rmin)-1),1):min(j1+(ceil(rmin)-1),nely)
                        e2 = (i2-1)*nely*nelz + (k2-1)*nely + j2;
                        k = k + 1;
                        iH(k) = e1;
                        jH(k) = e2;
                        sH(k) = max(0,rmin-sqrt((i1-i2)^2+(j1-j2)^2+(k1-k2)^2));
                    end
                 end
             end
         end
     end
end
Hk = sparse(iH,jH,sH);
Hs = sum(Hk,2);
\end{lstlisting}

\noindent and replace line~109, line~140 and line~149 (right part) by
\begin{lstlisting}[basicstyle=\footnotesize\ttfamily,breaklines=true,numbers=none,frame=tb,backgroundcolor=\color{beaublue}]
dVol = Hk*(dVol0./Hs);
\end{lstlisting}

\begin{lstlisting}[basicstyle=\footnotesize\ttfamily,breaklines=true,numbers=none,frame=tb,backgroundcolor=\color{beaublue}]
objsens = Hk*(objsens*normf./Hs);
\end{lstlisting}

\noindent and
\begin{lstlisting}[basicstyle=\footnotesize\ttfamily,breaklines=true,numbers=none,frame=tb,backgroundcolor=\color{beaublue}]
xphys = (Hk*xphys(:))./Hs; 
\end{lstlisting}

\noindent respectively.

\noindent (V) \textbf{MMA OPTIMIZATION PREPARATION AND INITIALIZATION:} Lines~102-110 provide this part of the code. Line~103 initializes design variable vector \texttt{x} and provides the unfiltered derivative of the volume constraint in vector \texttt{dVol0}. \texttt{x} is updated on line~104 using the active design variable vector \texttt{act}. \texttt{nMMA}, \texttt{mMMA}, \texttt{xphys}, \texttt{xMMA}, and \texttt{mvLt} indicate the number of design variables, number of the constraints, physical design vector, design variable used in the MMA, and external move limit for the MMA, respectively. These variables are defined on line~105. Vectors \texttt{xminvec} and \texttt{xmaxvec} define the minimum and maximum value of the design vector~(line~106). Lower (\texttt{low}) and upper (\texttt{upp}) values of the design vector are defined on line~107. The same line also initializes vectors \texttt{xold1} and \texttt{xold2}, which will be used to restore the old design vector during optimization. Other important parameters of the MMA, e.g., \texttt{cMMA}, \texttt{dMMA}, \texttt{a0} and \texttt{aMMA} are defined on line~108~\citep{svanberg1987}.  The code uses 2007 version of the MMA code\footnote{\url{https://www.smoptit.se/}}. As density filtering is volume preserving, we perform filtering of the derivatives of the volume constraint above the optimization loop on line~109 in vector \texttt{dVol}. \texttt{loop} records the optimization loop, and \texttt{change} tracks the absolute change in the design vector during optimization (line~110). 

\noindent (VI) \textbf{MMA OPTIMIZATION LOOP:} The MMA optimization loop contains five parts, which starts with \texttt{while} loop with termination conditions on \texttt{maxit} and \texttt{change}. Lines~111-158 describe this part of the code. Line~114 defines the definition of the \texttt{while} loop. \texttt{loop} is used to store the progress optimization's iterations line~113. 
\begin{itemize} 
	\setlength{\itemindent}{2em}\item [(VI.1)]\textbf{SOLVING THE FLOW BALANCE EQUATION} (Lines~114-122): This part of the code provides the pressure field in terms of the physical (design) variables. The flow coefficients ($K$, cf.~Eq.~\ref{Eq:Flowcoefficient}) and drainage coefficients ($D$, cf.~Eq.~\ref{Eq:stateequation})  of all FEs are recorded in vectors \texttt{Kc} (line~115) and \texttt{Dc} (line~116), respectively. The elemental flow coefficient matrix due to the Darcy law and drainage term, $\mathbf{A}_e$, (Eq.~\ref{Eq:FEAnumint}) is determined on line~117 and recorded in \texttt{Ae}. The global flow coefficient matrix, $\mathbf{A}$ (Eq.~\ref{Eq:PDEsolutionpressure}), (in lower triangular form) is determined on line~118  using \texttt{fsparse} and recorded in \texttt{AG}. Line~119 determines the flow coefficient matrix corresponding to the free pressure DOFs and records in \texttt{Aff}. The full matrix of \texttt{AG} is recovered on line~120. Line~121 determines pressure field, $\mathbf{p}$ (Eq.~\ref{Eq:PDEsolutionpressure}), and stores in \texttt{PF}. We use \texttt{decomposition} MATLAB function with `\texttt{ldl}' type and `\texttt{lower}' format for determining \texttt{PF}. Pressure load vector \texttt{PF}  is modified per the known pressure load conditions on line~122.
	
	\setlength{\itemindent}{2em}\item [(VI.1)]\textbf{DETERMINING CONSISTENT NODAL LOAD AND GLOBAL DISPLACEMENT VECTORS} (Lines~123-129): Line~124 provides global load vector \texttt{F} using matrix \texttt{TG} (line~79) and vector \texttt{PF} (line~122).  Young's modulus vector \texttt{E} is determined on line~125. The elemental stiffness matrix is stored in the vector form on line~126, which is further used for determining the global stiffness matrix $\mathbf{K}$ (Eq.~\ref{Eq:OPTI}) on line~127. The latter is stored in matrix \texttt{KG}. \texttt{Cholesky} factorization function \texttt{chol} is used on line~128. On line~129, the global displacement vector \texttt{U} is determined. On can also replace the direct solver with the multigrid-preconditioned CG~\citep{amir2014multigrid,amir2015revisiting} to obtain vectors \texttt{PF} and \texttt{U}.

	\setlength{\itemindent}{2em}\item [(VI.3)]\textbf{OBJECTIVE, CONSTRAINT AND THEIR SENSITIVITIES COMPUTATION} (Line~130-140): The objective is determined on line~131. $\bm{\lambda}_1$ is determined on line~132 and stored in \texttt{lam1}. Derivatives of the objective without load sensitivity terms are determined on line~133 and are stored in \texttt{objsT1}. The load sensitivities are determined on line~136 using \texttt{dC1k} (line~134, contribution from the Darcy law) and \texttt{dC1d} (line~135, contribution form the drainage term) and are stored in a vector \texttt{objesT2}. The final derivatives are determined on line~137 and are recorded in \texttt{objsens}. \texttt{lst} defines presence (\texttt{lst}=1) or absence (\texttt{lst} =0) of \texttt{objesT2} in vector \texttt{objsens}. On line~138, the volume fraction of the intermediate design is determined in \texttt{Vol}. Normalization parameter \texttt{normf} is defined and determined on line~139. \texttt{normf} is utilized to normalize the objective; thus, the objective derivatives consistently. The vector \texttt{objsens} is filtered while normalizing on line~140. 
	
	\setlength{\itemindent}{2em}\item [(VI.4)]\textbf{SETTING AND CALLING MMA OPTIMIZATION } (Lines~141-150): \texttt{xval} is initialized with \texttt{xMMA} on line~142. Vectors~\texttt{xminvec} and \texttt{xmaxvec} are updated on line~143 using vector \texttt{xval} and parameter \texttt{mvLt}. The MMA subroutine \texttt{mmasub} is called on line~144, which updates the new design variable \texttt{xmma}. Vectors \texttt{xold1} and \texttt{xold2} are updated on line~146. A new design vector \texttt{xnew} is determined using \texttt{xmma} on line~148. \texttt{change} is determined on line~147 using the new and old design vectors. Line~148 also updates \texttt{xMMA} using \texttt{xnew}. On line~149, the physical design vector \texttt{xphys} is updated, and filtering is performed. Next line updates \texttt{xphys} using information for the solid and void passive regions.
	\setlength{\itemindent}{2em}\item [(VI.5)]\textbf{PRINTING AND PLOTTING RESULTS} (Lines~151-157): We print \texttt{loop}, objective value, volume fraction, and \texttt{change} on line~152 using \texttt{fprintf} MATLAB function. MATLAB functions \texttt{cla}, \texttt{shiftdim}, \texttt{smooth3}, \texttt{patch}, \texttt{view}, \texttt{axis}, \texttt{drawnow} and \texttt{camlight}  are used to plot the optimized results~\citep{amir2014multigrid,amir2015revisiting} between lines~153-157.  On line~158, the \texttt{while} loop gets ended.
\end{itemize}
\begin{figure}
	\centering
	\includegraphics[scale =0.85]{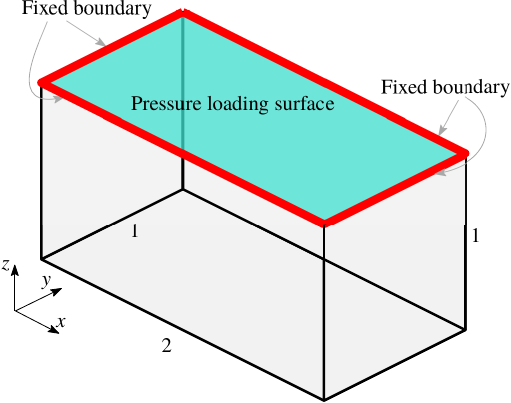}
	\caption{Design domain for a loadbearing lid structure. Pressure load is applied on the top surface, and all its edges are fixed.}
	\label{fig:lid}
\end{figure}
\begin{figure}
	\centering
	\begin{subfigure}{0.22\textwidth}
		\centering
		\includegraphics[scale=0.30]{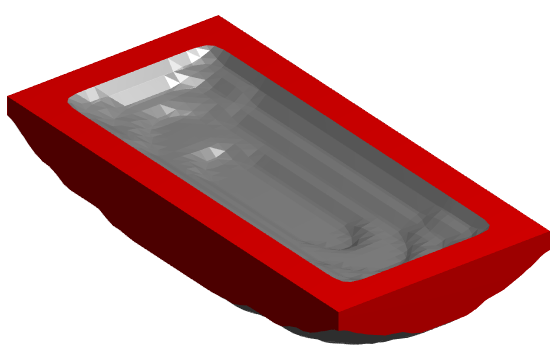}
		\caption{Isoparametric view}
		\label{fig:lidisoview}
	\end{subfigure}
	\quad
	\begin{subfigure}{0.22\textwidth}
		\centering
		\includegraphics[scale=0.30]{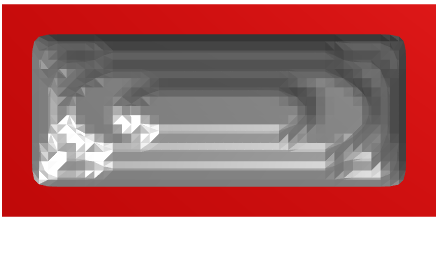}
		\caption{Top view}
		\label{fig:lidtop}
	\end{subfigure}
	\quad
	\begin{subfigure}{0.22\textwidth}
		\centering
		\includegraphics[scale=0.30]{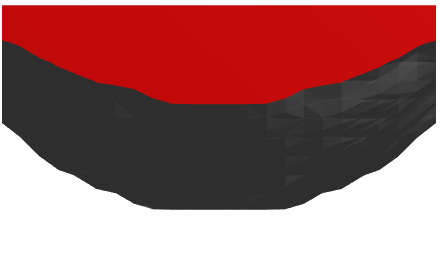}
		\caption{Side view}
		\label{fig:lidside}
	\end{subfigure}
	\quad
	\begin{subfigure}{0.22\textwidth}
		\centering
		\includegraphics[scale=0.35]{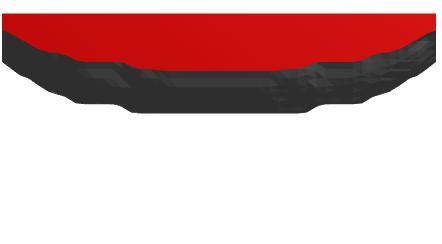}
		\caption{Front view}
		\label{fig:lidfront}
	\end{subfigure}
	\caption{Optimized pressure loadbearing lid structure in different views. The domain is parameterized using $48\times 24 \times 24$ FEs. The density value of the isosurface displayed is 0.3} \label{fig:lidsolution}
\end{figure}
\section{Results}\label{Sec:Sec4}
We provide four examples of pressure loadbearing structures to demonstrate the versatility and robustness of \texttt{TOPress3D}. In the material and flow definition, we set \texttt{E1} = 1,  \texttt{Emin} = $\texttt{E1}\times 10^{-5}$, \texttt{nu} = 0.3, \texttt{Kv} =1, \texttt{epsf} = $10^{-7}$, \texttt{r} =0.1, and \texttt{Dels} = 2. We use a 64-bit laptop with Processor Intel(R) Core(TM) i5-8265U \@ 1.60 GHz, RAM 8GB, Windows 11 Pro, and MATLAB 2022a for presenting the numerical results herein.
\subsection{Loadbearing Lid Structure}
The original state of \texttt{TOPress3D} is set for designing a pressure loadbearing lid structure~(Fig.~\ref{fig:lid}).  The problem is reported earlier in~\cite{du2004topological,sigmund2007topology,zhang2010topology,kumar2021topology}. 

The design domain, applied pressure load, and displacement boundary conditions are illustrated in Fig.~\ref{fig:lid}. The top surface of the domain experiences pressure load, whereas the bottom face receives zero pressure loading. All edges of the top surface are fixed. Dimension of the domain is considered to be $2\times1\times1$. While the problem has symmetry with respect to the vertical planes, we use the entire domain to obtain the optimized design to notice any deviation from the symmetry.

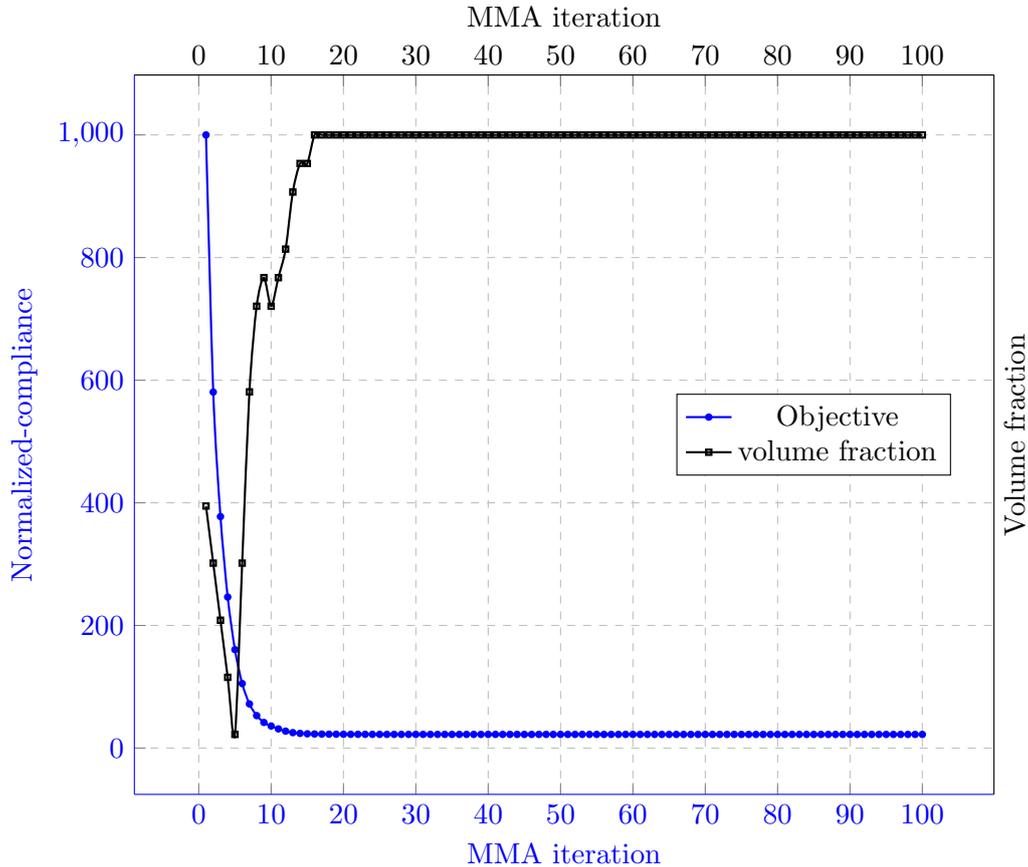
\begin{figure}[h!]
	\centering
	\begin{tikzpicture}
		\pgfplotsset{compat = 1.3}
		\begin{axis}[ blue,
			width = 0.60\textwidth,
			xlabel=MMA iteration,
			axis y line* = left,
			ylabel= Normalized-compliance,
			ymajorgrids=true,
			xmajorgrids=true,
			grid style=dashed ]
			\pgfplotstableread{lidobj.txt}\mydata;
			\addplot[smooth,blue,mark = *,mark size=1pt,style={thick}]
			table {\mydata};\label{plot1}
		\end{axis}
		\begin{axis}[
			width = 0.60\textwidth,
			axis y line* = right,
			ylabel= Volume fraction,
			axis x line*= top,
			xlabel=MMA iteration,
			ytick = {0.290,0.295,0.300},
			yticklabel style={/pgf/number format/.cd,fixed,precision=3},
			legend style={at={(0.95,0.5)},anchor=east}]
			\addlegendimage{/pgfplots/refstyle=plot1}\addlegendentry{Objective}
			\pgfplotstableread{lidvol.txt}\mydata;
			\addplot[smooth,black,mark = square,mark size=1pt,style={thick}]
			table {\mydata};
			\addlegendentry{volume fraction}
		\end{axis}
	\end{tikzpicture}
	\caption{Convergence plot for the loadbearing lid structure}
	\label{fig:Convergencelid}
\end{figure}

We call \texttt{TOPress3D} in the MATLAB command windows as
\begin{lstlisting}[basicstyle=\footnotesize\ttfamily,breaklines=true,numbers=none,frame=tb,backgroundcolor=\color{beaublue}]
TOPress3D(48,24,24,0.25,3,sqrt(3),0.20,10,1,100);
\end{lstlisting}
wherein \texttt{nelx} = 48, \texttt{nely} = 24, \texttt{nelz} = 24, \texttt{volf} = 0.25, \texttt{rmin} = $\sqrt 3$, \texttt{etaf} = 0.20, \texttt{betaf} = 10, \texttt{lst} = 1, \texttt{maxit} = 100.
The optimized loadbearing lid structure in different views is displayed in Fig.~\ref{fig:lidsolution}. The density value of the isosurface displayed is 0.3 for the optimized design (Fig.~\ref{fig:lidsolution}). Note that to obtain the top view, side view, and front view, one changes \texttt{view(3)} (line~159) to \texttt{view(90,90)}, \texttt{view(180,0)}, and \texttt{view(270,0)}, respectively. The obtained optimized design is symmetrical and provides a suitable chamber on the top surface to contain more fluid pressure while optimizing the strength of the design. The normalized-objective and volume fraction convergence histories are depicted in Fig.~\ref{fig:Convergencelid}. One notices that changes in the objective and volume fraction are insignificant from the 20$^\text{th}$ iteration onwards. The volume constraint remains active at the end of the optimization.

\begin{figure}
	\centering
	\includegraphics[scale =01]{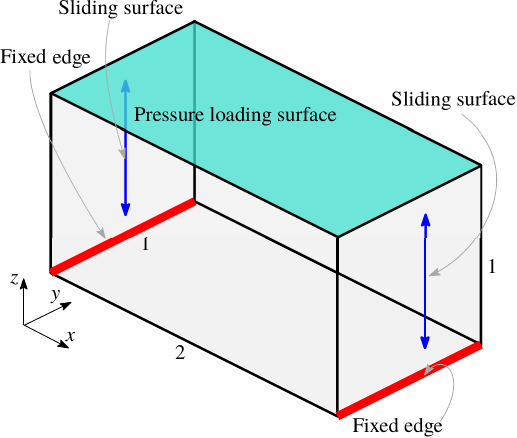}
	\caption{Design domain for an externally pressurized structure. The pressure load is applied on the top surface. The bottom, front and back faces get zero pressure loading. The left and right edges of the bottom face are fixed. The left and right faces slid in the vertical direction.}
	\label{fig:ExtPressS}
\end{figure}

\subsection{Externally Pressurized Structure}
An externally pressurized structure is optimized. This particular problem is previously reported in~\cite{kumar2021topology,zhang2010topology,du2004topological}. The design domain is depicted in Fig.~\ref{fig:ExtPressS}. The top surface of the structure experiences a fluidic pressure load, while the left and right edges of the bottom face are fixed, as depicted in the figure. The left and right faces can slide in the $z-$direction. The bottom face get zero pressure loading. The dimensions of the domain are considered to be $2\times1\times1$. Note that the optimization, in this case, is focused exclusively on the right symmetric half part with respect to the $yz$-plane of the domain.

\begin{figure}[h!]
	\begin{subfigure}{0.22\textwidth}
		\centering
		\includegraphics[scale=0.32]{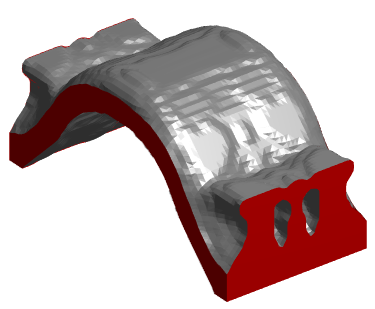}
		\caption{Isoparametric view}
		\label{fig:ExtPressSisoview}
	\end{subfigure}
	\quad
	\begin{subfigure}{0.22\textwidth}
		\centering
		\includegraphics[scale=0.35]{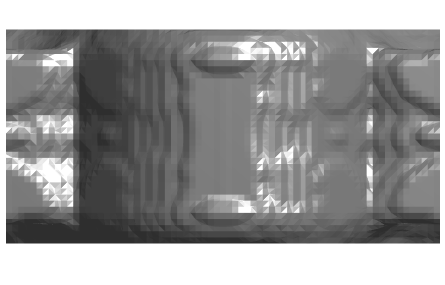}
		\caption{Top view}
		\label{fig:ExtPressStop}
	\end{subfigure}
	\quad \quad
	\begin{subfigure}{0.22\textwidth}
		\centering
		\includegraphics[scale=0.30]{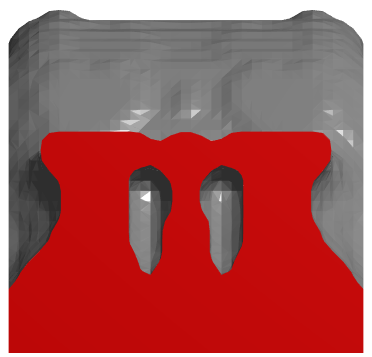}
		\caption{Side view}
		\label{fig:ExtPressSside}
	\end{subfigure}
	\quad
	\begin{subfigure}{0.22\textwidth}
		\centering
		\includegraphics[scale=0.35]{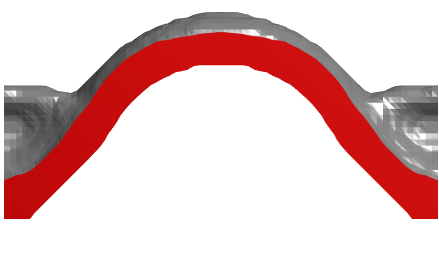}
		\caption{Front view}
		\label{fig:ExtPressSfront}
	\end{subfigure}
	\caption{Optimized externally pressurized structure in different views. The symmetrical half domain is parameterized using $36\times 36 \times 36$ FEs. The density value of the isosurface displayed is 0.3} \label{fig:ExtPressSsolution}
\end{figure}

For applying the pressure load, line~88 is modified to
\begin{lstlisting}[basicstyle=\footnotesize\ttfamily,breaklines=true,numbers=none,frame=tb,backgroundcolor=\color{beaublue}]
PF(BTface)= 0; PF(Tface) = Pin;
\end{lstlisting}
Boundary conditions are applied by replacing the lines~92-94 with the following code:
\begin{lstlisting}[basicstyle=\footnotesize\ttfamily,breaklines=true,numbers=none,frame=tb,backgroundcolor=\color{beaublue}]
fixnn = intersect(BTface,Rface);
fixedUdofs = [3*fixnn-2  3*fixnn-1  3*fixnn 3*Lface-2 3*Rface-1 3*Rface-2] ; 
\end{lstlisting}
In the plotting routine, one  replaces line~153 with the following code to plot the full optimized design from the symmetrical half result:
\begin{lstlisting}[basicstyle=\footnotesize\ttfamily,breaklines=true,numbers=none,frame=tb,backgroundcolor=\color{beaublue}]
cla;isovals = zeros(nelx*2,nely,nelz);
isovals(nelx+1:2*nelx,1:nely,1:nelz) = shiftdim(reshape(xphys,nely,nelz,nelx),2);
isovals(1:nelx,1:nely,1:nelz) = isovals(2*nelx:-1:nelx+1,1:nely,1:nelz);
\end{lstlisting}

\noindent With the above modifications \texttt{TOPress3D} code is called as
\begin{lstlisting}[basicstyle=\footnotesize\ttfamily,breaklines=true,numbers=none,frame=tb,backgroundcolor=\color{beaublue}]
TOPress3D(36,36,36,0.25,3,sqrt(3),0.2,10,1,100);
\end{lstlisting}
with \texttt{nelx} = 36, \texttt{nely} = 36, \texttt{nelz} = 36, \texttt{volf} = 0.25, \texttt{rmin} = $\sqrt 3$, \texttt{etaf} = 0.20, \texttt{betaf} = 10, \texttt{lst} = 1, \texttt{maxit} = 100. 

 The optimized design with different views is illustrated in Fig.~\ref{fig:ExtPressSsolution}. The results are displayed with isosurface 0.3. The symmetrical half-optimized design is suitably converted into a full optimized design using the above plotting code. We find that convergences for the objective and volume constraint are smooth, and the volume constraint remains active at the end of the optimization.
\subsection{Dam Structure}
\begin{figure}
	\centering
	\includegraphics[scale =1]{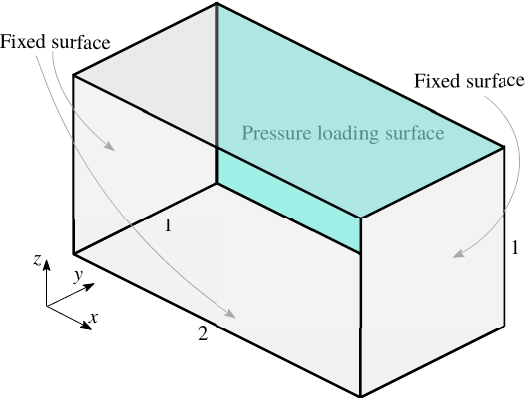}
	\caption{A Dam design domain. The bottom, left, and right faces are fixed. The back face experiences pressure load, whereas the front face receives no pressure load.}
	\label{fig:Dam}
\end{figure}

A dam structure, solved first in~\cite{sigmund2007topology}, is optimized herein. The design domain is shown in Fig.~\ref{fig:Dam}. The pressure load is applied on the back face of the domain, whereas the front face experiences zero pressure load. The left, right, and bottom faces are fixed. Note that designing an actual dam structure requires complicated loading and boundary conditions~\citep{sigmund2007topology}. The dimensions of the domain are $2\times1 \times1$. Utilizing the symmetry of the problem, only one-half of the domain is optimized.
\begin{figure}[h!]
	\begin{subfigure}{0.22\textwidth}
		\centering
		\includegraphics[scale=0.32]{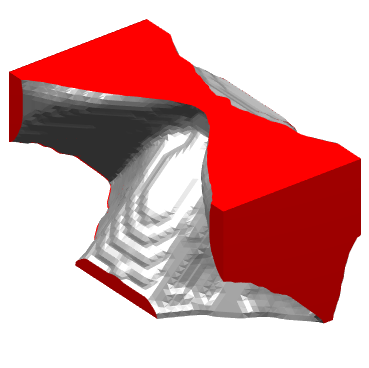}
		\caption{Isoparametric view}
		\label{fig:DamSisoview}
	\end{subfigure}
	\quad
	\begin{subfigure}{0.22\textwidth}
		\centering
		\includegraphics[scale=0.35]{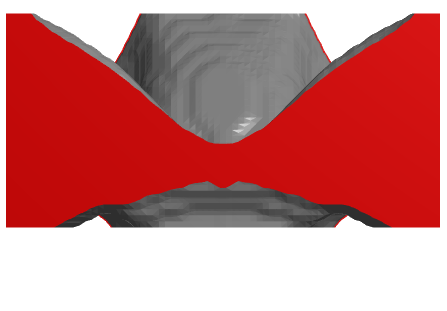}
		\caption{Top view}
		\label{fig:DamStop}
	\end{subfigure}
	\quad
	\begin{subfigure}{0.22\textwidth}
		\centering
		\includegraphics[scale=0.30]{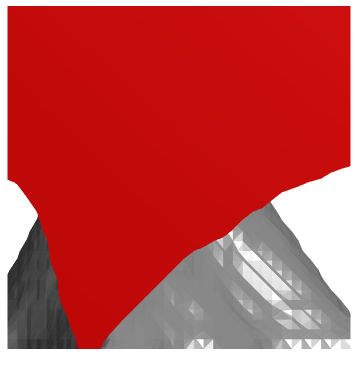}
		\caption{Side view}
		\label{fig:Damside}
	\end{subfigure}
	\quad
	\begin{subfigure}{0.22\textwidth}
		\centering
		\includegraphics[scale=0.35]{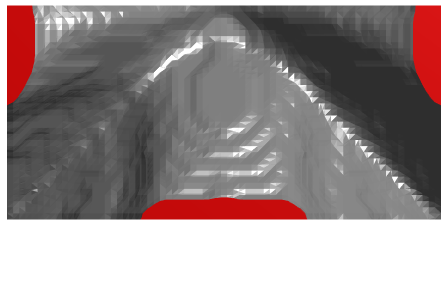}
		\caption{Front view}
		\label{fig:Damfront}
	\end{subfigure}
	\caption{Optimized pressure loadbearing dam structure in different views. The symmetrical half domain is parameterized using $36\times 36 \times 36$ FEs. The density value of the isosurface displayed is 0.3} \label{fig:DamSsolution}
\end{figure}

To solve this problem using \texttt{TOPress3D}, one modifies line~88 to
\begin{lstlisting}[basicstyle=\footnotesize\ttfamily,breaklines=true,numbers=none,frame=tb,backgroundcolor=\color{beaublue}]
PF(Fface) = 0; PF(Bface) = Pin;     
\end{lstlisting}
and line~92-94 to
\begin{lstlisting}[basicstyle=\footnotesize\ttfamily,breaklines=true,numbers=none,frame=tb,backgroundcolor=\color{beaublue}]
fixnn = unique([BTface,Rface]);
fixedUdofs = [3*fixnn-2  3*fixnn-1  3*fixnn  3*Lface-2];
\end{lstlisting}
In the plotting, we perform a similar code presented for an externally pressurized structure. Having done the above modification, the user can call \texttt{TOPress3D} code as
\begin{lstlisting}[basicstyle=\footnotesize\ttfamily,breaklines=true,numbers=none,frame=tb,backgroundcolor=\color{beaublue}]
TOPress3D(36,36,36,0.5,3,sqrt(3),0.2,10,1,100);
\end{lstlisting}
wherein \texttt{nelx} = 36, \texttt{nely} =36, \texttt{nelz} = 36, \texttt{volf} = 0.5, \texttt{rmin} = $\sqrt 3$, \texttt{etaf} = 0.20, \texttt{betaf} = 10, \texttt{lst} = 1, \texttt{maxit} = 100.

The optimized design in the full domain is shown in Fig.~\ref{fig:DamSsolution}. Different views are also depicted in the figure. The optimized design resembles that presented in~\cite{sigmund2007topology}. The convergence of optimization's progress is found to be smooth.  

\subsection{Externally Pressurized hull structure}
\begin{figure}
	\centering
	\includegraphics[scale =1]{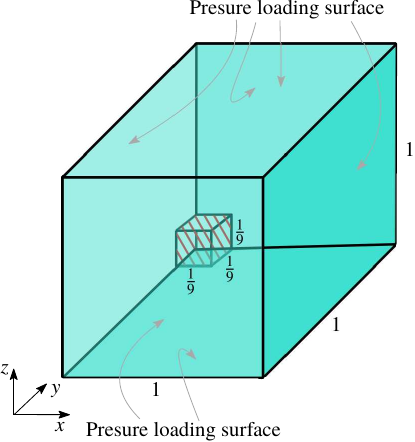}
	\caption{An externally pressurized hull structure. The structure is pressurized from all sides. A center cuboid void region is present, which is fixed.}
	\label{fig:hull}
\end{figure}

To demonstrate the code's capability with design domains having passive regions, we optimize a pressure hull structure herein~\citep{wang2020density}. Fig.~\ref{fig:hull} depicts the design domain of the structure. All the surfaces are pressurized from the outside. The hull's center contains a cuboid passive void region, which is fixed (Fig.~\ref{fig:hull}). The dimensions of the domain and void region are $1 \times 1 \times 1$ and $\frac{1}{9}\times \frac{1}{9} \times \frac{1}{9}$, respectively. Note that one may change the size of the void as per the requirement. 

One makes the following modifications in \texttt{TOPress3D} to solve this problem. Line~84 is replaced by
\begin{lstlisting}[basicstyle=\footnotesize\ttfamily,breaklines=true,numbers=none,frame=tb,backgroundcolor=\color{beaublue}]
elNrs = reshape(1:nel,nely, nelz, nelx);
v1 = elNrs(8*nely/18:10*nely/18,8*nelz/18:10*nelz/18,8*nelx/18:10*nelx/18);
[NDS, NDV] = deal(  [],[v1(:)] ); 
Vnode = unique(Pdofs(v1(:),:));
\end{lstlisting}
where \texttt{elNrs} matrix arranges the element number in 3D matrix fashion using \texttt{reshape} MATLAB inbuilt function. \texttt{v1} extracts the information of FEs which are inside the void region (Fig.~\ref{fig:hull}). Using \texttt{v1}, next, \texttt{NDV} is determined. Nodes constituting elements in \texttt{v1} are extracted in vector \texttt{Vnode} using matrix \texttt{Pdofs}. Now, to apply the pressure load, line~88 is changed to 
\begin{lstlisting}[basicstyle=\footnotesize\ttfamily,breaklines=true,numbers=none,frame=tb,backgroundcolor=\color{beaublue}]
PF(unique([BTface Bface Tface Fface Lface Rface])) = Pin;  PF(Vnode(:)) = 0;
\end{lstlisting}
and boundary conditions are applied by changing lines 92-94  to
\begin{lstlisting}[basicstyle=\footnotesize\ttfamily,breaklines=true,numbers=none,frame=tb,backgroundcolor=\color{beaublue}]
fixedUdofs = [ 3*Vnode(:)-2 3*Vnode(:)-1 3*Vnode(:)];   
\end{lstlisting}
After performing the above modification, \texttt{TOPress3D} code is called as
\begin{lstlisting}[basicstyle=\footnotesize\ttfamily,breaklines=true,numbers=none,frame=tb,backgroundcolor=\color{beaublue}]
TOPress3D(36,36,36,0.2,3,sqrt(3),0.2,10,1,100);
\end{lstlisting}
with \texttt{nelx} = 36, \texttt{nely} =36, \texttt{nelz} = 36, \texttt{volf} = 0.5, \texttt{rmin} = $\sqrt 3$, \texttt{etaf} = 0.20, \texttt{betaf} = 10, \texttt{lst} = 1, \texttt{maxit} = 100.
\begin{figure}
	\centering
	\includegraphics[scale =0.85]{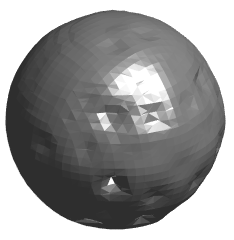}
	\caption{Optimized pressure loadbearing hull design.}
	\label{fig:hullsolution}
\end{figure}

The optimized pressure loadbearing hull structure is illustrated in Fig.~\ref{fig:hullsolution}. The optimized hull structure resembles that presented in~\cite{wang2020density}. Smooth and steady convergence characteristics are noticed for the objective and volume fraction.

The working and success of the presented \texttt{TOPress3D} are demonstrated above in four examples. We envision that the code will allow people new to the 3D topology optimization field to learn and explore different applications. The code can also be extended to solve pressure loadbearing structures with advanced constraints, e.g., stress constraints, buckling constraints, etc.

\section{Concluding remarks}\label{Sec:Sec5}
This paper introduces a MATLAB code, \texttt{TOPress3D}, comprising 158 lines, for 3D topology optimization for structures with design-dependent pressure loads. While such loads are encountered in various applications, addressing them within a TO framework proves challenging due to the dynamic nature of their magnitude, direction, and location during (especially at the beginning of) the optimization process. These challenges are particularly more evident in 3D TO problems, potentially posing difficulties for newcomers and students on the learning path. The developed \texttt{TOPress3D} can become a valuable tool and practical gateway for individuals entering the field, including newcomers, students, and researchers. The code utilizes the method of moving asymptotes to update design variables. Darcy's law with a drainage term is used to determine the pressure field in terms of the design values. The obtained pressure field is converted to consistent nodal loads. 

The code (Appendix~\ref{Sec:TOPress3D}) comprises six main subroutines, each explained in detail within the paper. Efficiency in matrix assembly is improved by representing mesh-related quantities as integers (MATLAB \texttt{int32}) and assembling only one half of the matrices. The transformation matrix is assembled before the optimization process, as its elemental part is independent of the design vector; which in turn saves computational time. To determine the state variables (pressure and displacement herein) one can also utilize the preconditioned iterative solvers as mentioned in \texttt{TOPress3D}.

The structure's compliance is minimized with specified volume constraints. The efficacy and robustness of the code are demonstrated for designing four pressure loadbearing structures. The original state of \texttt{TOPress3D} is  set for optimizing pressure loadbearing lid structure. The convergence plots for the objective and volume constraint are smooth. Extensions of the code are mentioned to  solve different pressure loadbearing structures. We anticipate that the code will serve as a valuable platform for learning, development, and extension to diverse applications involving design-dependent loads and opens several possibilities for future research.

\begin{appendices}
	
	\numberwithin{equation}{section}
	\section{Determining $\mathbf{\mathbf{K}}_\text{p}^e$, $\mathbf{K}^e_\text{Dp}$ and $\mathbf{T}_e$} \label{Sec:APPExpression}
	We provide the expressions for the flow matrices $\mathbf{\mathbf{K}}_\text{p}^e$ and $\mathbf{K}^e_\text{Dp}$, and  transformation matrix $\mathbf{T}_e$ for element~$e$. 
	
	As mentioned, hexahedral FEs are used to describe the design domain. Each FE contains 8 nodes (Fig.~\ref{fig:DOFs}). To interpolate the geometry, displacement field and pressure field, the Lagrange shape functions, noted below for the $k-\text{th}$ node, are used~\citep{zienkiewicz2005finite}
	\begin{equation}
		N_k = \frac{1}{8}(1+\xi_k\xi)(1+\eta_k\eta)(1+\zeta_k\zeta);\, k = 1\, \text{to}\, 8
	\end{equation}
	where $(\xi_k,\,\eta_k,\,\zeta_k)$ represents the coordinates of node~$k$ of the element in the  $(\xi,\,\eta,\,\zeta)$ system (Fig.~\ref{fig:DOFs}). We note the following expressions of  $\mathbf{\mathbf{K}}_\text{p}^e$, $\mathbf{K}^e_\text{Dp}$ and $\mathbf{T}_e$ for element~$e$ from Sec.~\ref{Sec:Sec2} as
	\begin{equation}\label{Eq:Appen_expressions}
		\begin{aligned}
			\mathbf{\mathbf{K}}_\text{p}^e &= \int_{\Omega_e} K~ \mathbf{B}^\top_\text{p} \mathbf{B}_\text{p} \text{d} V \\&= \int_{-1}^{+1}\int_{-1}^{+1}\int_{-1}^{+1} \left( K~ \mathbf{B}^\top_\text{p} \mathbf{B}_\text{p}\right) |\det \mathbf{J}| d\xi d\eta d\zeta \\
			\mathbf{K}^e_\text{Dp} &=  \int_{\Omega_e}D ~\mathbf{N}^\top_\text{p} \mathbf{N}_\text{p} \text{d} V \\&= \int_{-1}^{+1}\int_{-1}^{+1}\int_{-1}^{+1} \left( D ~\mathbf{N}^\top_\text{p} \mathbf{N}_\text{p}\right) |\det \mathbf{J}| d\xi d\eta d\zeta \\
			\mathbf{T}_i &=-\int_{\mathrm{\Omega}_e} \trr{\mathbf{N}}_\mathbf{u} \mathbf{B}_\text{p}  \text{d} V\\& =-\int_{-1}^{+1}\int_{-1}^{+1}\int_{-1}^{+1} \trr{\mathbf{N}}_\mathbf{u} \mathbf{B}_\text{p}  |\det \mathbf{J}| d\xi d\eta d\zeta 
		\end{aligned}
	\end{equation}
	where we have used $dV = |\det \mathbf{J}| d\xi d\eta d\zeta $ and $\mathbf{J}$ is the $(3\times 3)$ Jacobian matrix. The integration is preformed numerically using the Gauss quadrature~\citep{zienkiewicz2005finite}, that yields\footnote{The symbolic calculation is performed first, then using the coordinates of a unit cube; these matrices are determined in numerical forms.}
	\begin{equation}
		{\mathbf{K}}_\text{p}^e|_{K=1}=	\frac{1}{12}
		\begin{bmatrix}
			4 &  &  &  &  &  &  &  \\
			0 & 4 &  &  &  &  &  &  \\
			-1 & 0 & 4 &  &  &  &  &  \\
			0 & -1 & 0 & 4 &  &  & \text{Sym.} &  \\
			0 & -1 & -1 & -1 & 4 &  &  &  \\
			-1 & 0 & -1 & -1 & 0 & 4 &  &  \\
			-1 & -1 & 0 & -1 & -1 & 0 & 4 &  \\
			-1 & -1 & -1 & 0 & 0 &-1 & 0 & 4 
		\end{bmatrix} 
	\end{equation}

	\begin{equation}
		\setlength{\arraycolsep}{4pt}
		{\mathbf{K}}_\text{Dp}^e|_{D=1}=	\frac{1}{216}
		\begin{bmatrix}
			8 &  &  &  &  &  &  &  \\
			4 & 8 &  &  &  &  &  &  \\
			2 & 4 & 8 &  &  &  &  &  \\
			4 & 2 & 4 & 8 &  &  & \text{Sym.} &  \\
			4 & 2 & 1 & 2 & 8 &  &  &  \\
			2 & 4 & 2 & 1 & 4 & 8 &  &  \\
			1 & 2 & 4 & 2 & 2 & 4 & 8 &  \\
			2 & 1 & 2 &4 & 4&2 & 4 & 8 
		\end{bmatrix} 
	\end{equation}

	\begin{equation}
		\setlength{\arraycolsep}{4pt}
		\mathbf{T}_e=\frac{1}{72}
		\begin{bmatrix}
			-4 & 4 & 2 &-2 &-2 & 2 & 1 &-1\\
			-4 &-2 & 2 & 4 &-2 &-1 & 1 & 2\\
			-4 &-2 &-1 &-2 & 4  &2  &1 & 2\\
			-4  &4  &2 &-2 &-2  &2 & 1 &-1\\
			-2 &-4  &4 & 2 &-1 &-2 & 2 & 1\\
			-2 &-4 &-2 &-1 & 2 & 4 & 2 & 1\\
			-2 & 2 & 4 &-4 &-1  &1 & 2 &-2\\
			-2 &-4 & 4 & 2 &-1 &-2 & 2 & 1\\
			-1& -2 &-4 &-2 & 1 & 2 & 4 & 2\\
			-2 & 2 & 4 &-4 &-1 & 1 & 2 &-2\\
			-4 &-2 & 2 & 4 &-2 &-1 & 1 & 2\\
			-2 &-1 &-2 &-4 & 2 & 1 & 2 & 4\\
			-2 & 2 & 1 &-1 &-4 & 4 & 2 &-2\\
			-2 &-1 & 1 & 2 &-4 &-2 & 2 & 4\\
			-4 &-2 &-1 &-2 & 4 & 2 & 1 & 2\\
			-2 & 2  &1 &-1 &-4 & 4 & 2 &-2\\
			-1 &-2 & 2 & 1 &-2 &-4 & 4 & 2\\
			-2 &-4 &-2 &-1 & 2 & 4 & 2 & 1\\
			-1 & 1 & 2 &-2 &-2 & 2 & 4 &-4\\
			-1 &-2  &2 & 1 &-2 &-4 & 4 & 2\\
			-1 &-2 &-4 &-2 & 1 & 2 & 4 & 2\\
			-1&  1  &2 &-2 &-2 & 2 & 4 &-4\\
			-2 &-1 & 1 & 2 &-4 &-2 & 2 & 4\\
			-2& -1 &-2 &-4 & 2 & 1 & 2  &4
		\end{bmatrix} 
	\end{equation}
\onecolumn
\section{The MATLAB code: \texttt{TOPress3D}}\label{Sec:TOPress3D}
\begin{lstlisting}[basicstyle=\footnotesize\ttfamily,breaklines=true,backgroundcolor=\color{beaublue}]
function TOPress3D(nelx,nely,nelz,volfrac,penal,rmin,etaf,betaf,lst,maxit)
%% ___PART 1.____________________________MATERIAL AND FLOW PARAMETERS
E1 = 1;                                              % Youngs' Modulus of solid
Emin = E1*1e-5;                                      % Youngs's Modulus of void
nu = 0.30;                                           % Poisson's ratio
[Kv,epsf,r,Dels] = deal(1,1e-7,0.1,2);               % Flow parameters
[Ds, Kvs]= deal((log(r)/Dels)^2*epsf,Kv*(1 - epsf));    % Flow parameters 
%% ____PART 2._________FINITE ELEMENT ANALYSIS PREPARATION and NON-DESIGN REGIONS
[ndx,ndy,ndz] = deal(nelx+1,nely+1,nelz+1);          % Node number in x, y and z
[nel,nno] = deal(nelx*nely*nelz, ndx*ndy*ndz);       % Number of elements and nodes
nodenrs = int32(reshape( 1 : nno,ndy, ndz, ndx ) );   
edofVec = reshape( 3 * nodenrs( 1 : nely, 1 : nelz, 1 : nelx ) + 1, nel, 1);
Udofs = edofVec + int32( [0,1,2,3*ndy*ndz+[0,1,2,-3,-2,-1],-3,-2,-1,3*ndy+...
[0,1,2],3*ndy*(ndz+1)+[0,1,2,-3,-2,-1],3*ndy+[-3,-2,-1]]); % Displacement DOFs matrix
[Pdofs,allPdofs, allUdofs] = deal(Udofs(:,3:3:end)/3,1:nno,1:3*nno); % Press DOFs, all DOFs disp DOFs
Tface = (ndy*nelz+1:ndz*ndy:nno-nely)+ (0:nely)'; 
BTface=(1:ndz*ndy:nno-nely)+ (0:nely)';
[Tface, BTface] = deal(Tface(:)', BTface(:)');
[Lface,Rface] = deal((1:ndy*ndz), (ndy*ndz*nelx+1:nno)); % Det left and right faces
[Bface,Fface] = deal((ndy:ndy:nno),(1:ndy:nno-nely));  % Det front and back faces
[skI, skII, spI, spII] = deal( [ ] );                                       
for j = 1 : 24
skI = cat( 2, skI, j : 24 );
skII = cat( 2, skII, repmat( j, 1, 24 - j + 1 ) );
end
[iK, jK ] = deal( Udofs( :,  skI )', Udofs( :, skII )' );
Iar = sort( [ iK( : ), jK( : ) ], 2, 'descend' ); clear iK jK % Reduced assembly index
Ke = 1/(1+nu)/(2*nu-1)/144 *( [ -32;-6;-6;8;6;6;10;6;3;-4;-6;-3;-4;-3;-6;10;...
3;6;8;3;3;4;-3;-3; -32;-6;-6;-4;-3;6;10;3;6;8;6;-3;-4;-6;-3;4;-3;3;8;3;...
3;10;6;-32;-6;-3;-4;-3;-3;4;-3;-6;-4;6;6;8;6;3;10;3;3;8;3;6;10;-32;6;6;...
-4;6;3;10;-6;-3;10;-3;-6;-4;3;6;4;3;3;8;-3;-3;-32;-6;-6;8;6;-6;10;3;3;4;...
-3;3;-4;-6;-3;10;6;-3;8;3;-32;3;-6;-4;3;-3;4;-6;3;10;-6;6;8;-3;6;10;-3;...
3;8;-32;-6;6;8;6;-6;8;3;-3;4;-3;3;-4;-3;6;10;3;-6;-32;6;-6;-4;3;3;8;-3;...
3;10;-6;-3;-4;6;-3;4;3;-32;6;3;-4;-3;-3;8;-3;-6;10;-6;-6;8;-6;-3;10;-32;...
6;-6;4;3;-3;8;-3;3;10;-3;6;-4;3;-6;-32;6;-3;10;-6;-3;8;-3;3;4;3;3;-4;6;...
-32;3;-6;10;3;-3;8;6;-3;10;6;-6;8;-32;-6;6;8;6;-6;10;6;-3;-4;-6;3;-32;6;...
-6;-4;3;6;10;-3;6;8;-6;-32;6;3;-4;3;3;4;3;6;-4;-32;6;-6;-4;6;-3;10;-6;3;...
-32;6;-6;8;-6;-6;10;-3;-32;-3;6;-4;-3;3;4;-32;-6;-6;8;6;6;-32;-6;-6;-4;...
-3;-32;-6;-3;-4;-32;6;6;-32;-6;-32]+nu*[ 48;0;0;0;-24;-24;-12;0;-12;0;...
24;0;0;0;24;-12;-12;0;-12;0;0;-12;12;12;48;0;24;0;0;0;-12;-12;-24;0;-24;...
0;0;24;12;-12;12;0;-12;0;-12;-12;0;48;24;0;0;12;12;-12;0;24;0;-24;-24;0;...
0;-12;-12;0;0;-12;-12;0;-12;48;0;0;0;-24;0;-12;0;12;-12;12;0;0;0;-24;...
-12;-12;-12;-12;0;0;48;0;24;0;-24;0;-12;-12;-12;-12;12;0;0;24;12;-12;0;...
0;-12;0;48;0;24;0;-12;12;-12;0;-12;-12;24;-24;0;12;0;-12;0;0;-12;48;0;0;...
0;-24;24;-12;0;0;-12;12;-12;0;0;-24;-12;-12;0;48;0;24;0;0;0;-12;0;-12;...
-12;0;0;0;-24;12;-12;-12;48;-24;0;0;0;0;-12;12;0;-12;24;24;0;0;12;-12;...
48;0;0;-12;-12;12;-12;0;0;-12;12;0;0;0;24;48;0;12;-12;0;0;-12;0;-12;-12;...
-12;0;0;-24;48;-12;0;-12;0;0;-12;0;12;-12;-24;24;0;48;0;0;0;-24;24;-12;...
0;12;0;24;0;48;0;24;0;0;0;-12;12;-24;0;24;48;-24;0;0;-12;-12;-12;0;-24;...
0;48;0;0;0;-24;0;-12;0;-12;48;0;24;0;24;0;-12;12;48;0;-24;0;12;-12;-12;...
48;0;0;0;-24;-24;48;0;24;0;0;48;24;0;0;48;0;0;48;0;48 ] ); % Elem stiffness matrix  
Ke0( tril( ones( 24 ) ) == 1 ) = Ke';
Ke0 = reshape( Ke0, 24, 24 ); 
Ke0 = Ke0 + Ke0' - diag( diag( Ke0 ) );     % Extracting full form
for j = 1 : 8
spI = cat( 2, spI, j : 8 );
spII = cat( 2, spII, repmat( j, 1, 8-j + 1 ) );
end
[iP, jP ] = deal( Pdofs( :,  spI )', Pdofs( :, spII )' );
IarP = sort( [ iP( : ), jP( : ) ], 2, 'descend' ); clear iP jP   
Kpl = [4;0;-1;0;0;-1;-1;-1;4;0;-1;-1;0;-1;-1;4;0;-1;-1;0;-1;4;-1;-1;-1; ...
0;4;0;-1;0;4;0;-1;4;0;4]/12;                         % Flow matrix due to Darcy
KDpl = [8;4;2;4;4;2;1;2;8;4;2;2;4;2;1;8;4;1;2;4;2;8;2;1;2;4;8;4;2;4;8;...
4;2;8;4;8]/216;                              % Flow matrix due to drainage term
[Kp(tril( ones(8))== 1), KDp(tril(ones(8))==1)]= deal(Kpl',KDpl);
[Kp,KDp]= deal(reshape( Kp, 8, 8 ),reshape( KDp, 8, 8 ));
Kp = Kp + Kp' - diag( diag( Kp ) ); KDp = KDp + KDp' - diag( diag( KDp ) );
[lKe, lKpl] = deal(length(Ke), length(Kpl));
Te = [-4;-4;-4;-4;-2;-2;-2;	-2;	-1;	-2;-4;-2;-2;-2;-4;-2;-1;-2;-1;-1;-1;-1;...
-2;-2;4;-2;-2;4;-4;-4;2;-4;-2;2;-2;-1;2;-1;-2;2;-2;-4;1;-2;-2;1;-1;-1;2;...
2;-1;2;4;-2;4;4;-4;4;2;-2;1;1;-1;1;2;-2;2;2;-4;2;1;-2;-2;4;-2;-2;2;-1;-4;...
2;-2;-4;4;-4;-1;2;-2;-1;1;-1;-2;1;-2;-2;2;-4;-2;-2;4;-2;-1;2;-1;-1;1;-1;...
-2;2;-4;-4;4;-4;-2;2;-2;-2;1;-2;-4;2;2;-1;2;2;-2;4;1;-2;2;1;-1;1;4;-2;2;...
4;-4;4;2;-4;2;2;-2;1;1;1;1;1;2;2;2;2;4;2;1;2;2;2;1;2;4;2;4;4;4;4;2;2;-1;...
2;2;-1;1;1;-2;1;2;-2;2;4;-2;4;2;-2;2;1;-4;2;2;-4;4;4]/72;
iT = reshape(kron(Udofs,int32(ones(8,1)))',192*nel,1);
jT = reshape(kron(Pdofs,int32(ones(1,24)))',192*nel,1);
Ts = reshape(Te(:)*ones(1,nel), 192*nel, 1);        % Elemental transformation matrix
TG = fsparse(iT, jT, Ts); clear Te iT jT Ts         % Global transformation matrix
IFprj=@(xv,etaf,betaf)((tanh(betaf*etaf) + tanh(betaf*(xv-etaf)))/... % Proj fun
(tanh(betaf*etaf) + tanh(betaf*(1 - etaf))));
dIFprj=@(xv,etaf,betaf) betaf*(1-tanh(betaf*(xv-etaf)).^2)...
/(tanh(betaf*etaf)+tanh(betaf*(1-etaf))); % Derivative of the projection function
[NDS, NDV ] = deal( [], [] );
act = setdiff((1 : nel)', union( NDS, NDV ));
%% ____PART 3.______PRESSURE & STRUCTURE B.C's, LOADs
[PF, Pin] =deal(0.00001*ones(nno,1),1);        % Pressure-field preparation
PF(BTface) = 0; PF(Tface) = Pin;               % Applying pressure load
fixedPdofs = allPdofs(PF~=0.00001);            % Given P-dofs
freePdofs  = setdiff(allPdofs,fixedPdofs);     % Free P-dofs
pfixeddofsv = [fixedPdofs' PF(fixedPdofs)];    % p-fixed and its value
fixnn = unique([intersect(Tface,Lface),intersect(Tface,Rface),...
intersect(Tface,Fface),intersect(Tface,Bface)]);
fixedUdofs = [3*fixnn-2  3*fixnn-1  3*fixnn ];    % Fixed displ.
freeUdofs = setdiff(allUdofs,fixedUdofs);         % Free dofs for displ.
[U, lam1] = deal(zeros(3*nno,1),zeros(nno,1));    % lam1:Lagrange mult.
%% ___PART 4._________________________________________FILTER PREPARATION
[dy,dz,dx]=meshgrid(-ceil(rmin)+1:ceil(rmin)-1,...
-ceil(rmin)+1:ceil(rmin)-1,-ceil(rmin)+1:ceil(rmin)-1 );
h = max( 0, rmin - sqrt( dx.^2 + dy.^2 + dz.^2 ) );      % Conv. kernel                
Hs = imfilter( ones( nely, nelz, nelx ), h);             % Matrix of weights (filter)  
%% ___PART 5.__________________________MMA OPTIMIZATION PREPARATION & INITIALIZATION
[x,dVol0] = deal(zeros(nel,1),ones(nel,1)/(nel*volfrac)); % Design var. vol. cont. der
x(act) = (volfrac*(nel-length(NDV))-length(NDS) )/length(act); x(NDS) = 1;  % Updating
[nMMA,mMMA,xphys,xMMA,mvLt] = deal(length(act),1,x,x(act),0.1);  % Different variables
[xminvec,xmaxvec] = deal(zeros(nMMA,1),ones(nMMA,1));      % Min. & Max vector for MMA
[low, upp, xold1,xold2] = deal(xminvec,xmaxvec,xMMA,xMMA);    % Low and Upp limits MMA
[cMMA,dMMA, a0, aMMA] = deal(1000*ones(mMMA,1),zeros(mMMA,1),1,zeros(mMMA,1));                                       
dVol = imfilter(reshape(dVol0, nely, nelz, nelx)./Hs,h); % Filtered volume sensitivity
[loop, change] =deal(0,1);   
%% ____PART 6._____________________________________MMA OPTIMIZATION LOOP
while(loop<maxit && change>0.0001)
loop = loop + 1;                                       % Updating the opt. iteration
%___PART 6.1__________SOLVING FLOW BALANCE EQUATION
Kc = Kv*(1-(1-epsf)*IFprj(xphys,etaf,betaf));                     % Flow coefficient
Dc = Ds*IFprj(xphys,etaf,betaf);                              % Drainage coefficient
Ae = reshape(Kpl(:)*Kc' + KDpl(:)*Dc',lKpl*nel,1);           % Elemental flow matrix
AG = fsparse(IarP(:,1),IarP(:,2),Ae,[nno, nno]) ;               % Global flow matrix
Aff = AG(freePdofs,freePdofs);                           % AG for free pressure dofs
AG = AG + AG' - diag( diag( AG ) ) ;                           % Determining full AG
PF(freePdofs) = decomposition(Aff,'ldl','lower')\(-AG(freePdofs,fixedPdofs)*pfixeddofsv(:,2)); % Solving for pressure
PF(pfixeddofsv(:,1)) = pfixeddofsv(:,2);                             % Final P-field
%__PART 6.2_DETERMINING CONSISTENT NODAL LOADS and GLOBAL Disp. Vector
F = -TG*PF;                                                 % Dertmining nodal forces
E = Emin + xphys.^penal*(E1 - Emin);                         % Material interpolation
Ks = reshape(Ke(:)*E',lKe*nel,1);                        % Elemental stiffness matrix
KG = fsparse( Iar( :, 1 ), Iar( :, 2 ), Ks, [3*nno, 3*nno ] ); % Global stiff matrix
L = chol( KG( freeUdofs, freeUdofs ), 'lower' );
U(freeUdofs ) = L' \ ( L \ F( freeUdofs ) );   
%__PART 6.3__OBJECTIVE, CONSTRAINT and THEIR SENSITIVITIES COMPUTATION
obj = U'*F;                                                 % Determining objective
lam1(freePdofs) = (2*U(freeUdofs)'*TG(freeUdofs,freePdofs))/Aff;   % Lagrange mult.
objsT1 = -(E1 - Emin)*penal*xphys.^(penal - 1).*sum(([U(Udofs)]*Ke0).*[U(Udofs)],2);
dC1k = -dIFprj(xphys,etaf,betaf).* sum((lam1(Pdofs)*(Kvs*Kp)) .* PF(Pdofs),2);
dC1d =  dIFprj(xphys,etaf,betaf).* sum((lam1(Pdofs)*(Ds*KDp)) .* PF(Pdofs),2);
objsT2 = dC1k + dC1d; 
objsens = (objsT1 + lst*objsT2);                              % Final sensitivities
Vol = sum(xphys)/(nel*volfrac)-1;                             % Volume fraction
if(loop ==1), normf = 1000/(obj);end
objsens = imfilter(reshape(objsens*normf, nely, nelz, nelx)./Hs,h); % Obj. sens.
%___PART 6.4______________________SETTING and CALLING MMA OPTIMIZATION
xval = xMMA;
[xminvec, xmaxvec]= deal(max(0, xval - mvLt),min(1, xval + mvLt));
[xmma,~,~,~,~,~,~,~,~,low,upp] = mmasub(mMMA,nMMA,loop,xval,xminvec,xmaxvec,...
xold1,xold2,obj*normf,objsens(act),Vol,dVol(act)',low,upp,a0,aMMA,cMMA,dMMA); 
[xold2,xold1,xnew] = deal(xold1, xval, xmma);    % Updating
change = max(abs(xnew-xMMA)); xMMA = xnew;   % Calculating chan and updating solu
xphys(act) = xnew; 
xphys = imfilter(reshape(xphys, nely, nelz, nelx),h)./Hs;% Filering the Phy. vector
xphys = xphys(:); xphys(NDS)= 1; xphys(NDV)= 0; % Upd xphys for active/passive region
%___PART 6.5_____________________________PRINTING and PLOTTING RESULTS
fprintf(' It.:%5i Obj.:%11.4f Vol.:%7.3f ch.:%7.3f\n',loop,obj*normf,mean(xphys),change);
cla; isovals = shiftdim(reshape( xphys, nely, nelz, nelx ), 2 );
isovals = smooth3( isovals, 'box', 1 );
patch(isosurface(isovals, 0.30),'FaceColor',[0.6 0.6 0.6],'EdgeColor','none');
patch(isocaps(isovals, 0.30),'FaceColor','r','EdgeColor','none');
view( 3 ); axis equal tight off; drawnow, camlight; 
end
\end{lstlisting}
	\lstdefinestyle{nonumbers}
{numbers=none}
\begin{lstlisting}[style=nonumbers]
	%%%%%%%%%%%%%%%%%%%%%%%%%%%%%%%%%%%%%%%%%%%%%%%%%%%%%%%%%%%%%%%%%%%%%%%%%%%%%%%%%%%%%%
	%    The code, TOPress3D,  is provided for  pedagogical purposes. A  detailed        %
	%    description is presented in the paper:"TOPress3D: 3D topology optimization      %       
	%    with design-dependent pressure loads in MATLAB"   Optimization and Engineering, % 
	%     2024.                                                                          %
	%    One can download the code and its extensions for the different problems         %
	%    from the online supplementary material and also from:                           %
	%                                      https://github.com/PrabhatIn/TOPress3D        %
	%    Please send your comment to: pkumar@mae.iith.ac.in                              %
	%    One may also refer to the following four papers:                                % 
	%                                                                                    %
	%    1. Kumar P, Frouws JS, Langelaar M (2020) Topology optimization of fluidic      %
	%    pressure-loaded structures and compliant mechanisms using the Darcy method.     %
	%    Structural and Multidisciplinary Optimization 61(4):1637-1655                   %
	%    2. Kumar P, Langelaar M (2021) On topology optimization of design-dependent     % 
	%    pressure-loaded three-dimensional structures and compliant mechanisms.          %
	%    International Journal for Numerical Methods in Engineering 122(9):2205-2220     %
	%    3. Kumar P. (2023) TOPress: a MATLAB implementation for topology optimization   %
	%    of structures subjected to design-dependent pressure loads.                     % 
	%    Structural and Multidisciplinary Optimization 66(4):97                          %
	%    4. Kumar P. (2023) SoRoTop: a hitchhiker's guide to topology optimization       %
	%    MATLAB code for design-dependent pneumatic-driven soft robots                   % 
	%    Optimization and Engineering: 2024                                              %
	%                                                                                    %
	%    Disclaimer:                                                                     %
	%    The author does not guarantee that the code is free from erros but reserves     %
	%    all rights. Further, the author shall not be liable in any event caused by      % 
	%    use of the above mentioned code and its extensions                              %
	%                                                                                    %
	%%%%%%%%%%%%%%%%%%%%%%%%%%%%%%%%%%%%%%%%%%%%%%%%%%%%%%%%%%%%%%%%%%%%%%%%%%%%%%%%%%%%%%
\end{lstlisting}
\end{appendices}




\end{document}